\pdfoutput=1

\documentclass[
a4paper, 
titlepage, 
bibliography=totoc, 
12pt, 
BCOR17mm, 
DIV12, 
headinclude, 
footinclude=false]
{scrbook}

\usepackage{graphicx}				
\usepackage{pdfpages}
\usepackage{here}
\usepackage{epstopdf}

\usepackage[english]{babel}
\usepackage[utf8]{inputenc}
\usepackage[T1]{fontenc}

\usepackage{fancyhdr}
\usepackage{setspace}

\usepackage{xcolor}

\usepackage{amsmath}
\usepackage{amsfonts}
\usepackage{amssymb}
\usepackage{hyperref}
\hypersetup{
	colorlinks,
	citecolor=blue,
	filecolor=black,
	linkcolor=black,
	urlcolor=black
}
\usepackage[square,comma,super]{natbib}
\usepackage{comment}
\usepackage{slashed}
\usepackage{ytableau}
\usepackage{wasysym}
\usepackage{geometry}
\usepackage[hang,flushmargin]{footmisc}
\usepackage{setspace}
\ytableausetup{smalltableaux}

\addto\captionsngerman{}

\definecolor{indigoblue}{cmyk}{1, .77, .34, .21}

\newcommand{\LMUTitle}[9]{
  \thispagestyle{empty}				
  \vspace*{\stretch{1}}				
  {\parindent0cm\rule{\linewidth}{.7ex}}  			
  \vspace*{\stretch{1}}	
  {\parindent0cm\rule{\linewidth}{.3ex}}  
     
  \begin{flushright}
    \vspace*{\stretch{1}}
    \bfseries\scshape\Huge #1\\
    \vspace*{\stretch{1}}
    \sffamily\bfseries\large #2\\
    \vspace*{\stretch{1}}
  \end{flushright}
  	
  {\parindent0cm\rule{\linewidth}{.3ex}}
  \vspace*{\stretch{1}} 
  {\parindent0cm\rule{\linewidth}{.7ex}}
  \vspace*{\stretch{5}}

  \begin{center}
    \includegraphics[width=10cm]{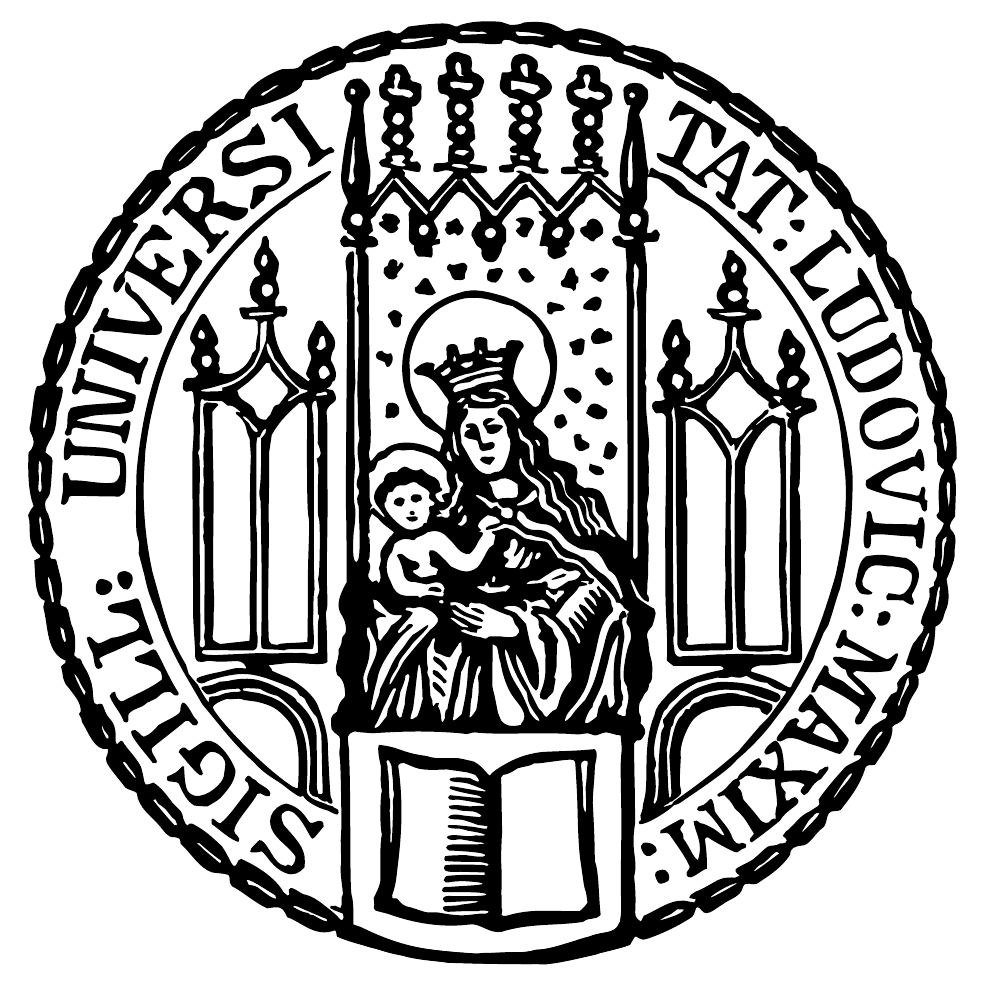}
  \end{center}
  
  \vspace*{\stretch{1}}
  \begin{center}\sffamily\LARGE{#5}
  \end{center}

  \newpage								
  \thispagestyle{empty}

  \cleardoublepage
  \thispagestyle{empty}

  \vspace*{\stretch{1}}
  {\parindent0cm
  \rule{\linewidth}{.7ex}}
  \begin{center}
    \vspace*{\stretch{1}}
    \bfseries\Huge Interactions in MacDowell-Mansouri Gravitation\\
    \vspace*{\stretch{1}}
  \end{center}
  \rule{\linewidth}{.7ex}

  \vspace*{\stretch{4}}
  \begin{center}
    \Large\textbf{Bachelor thesis}\\
    \Large Faculty of physics\\
    \Large Ludwig--Maximilians--University\\
    \Large Munich\\
    \vspace*{\stretch{1}}
    \Large submitted by\\
    \Large\textbf{#2}\\
    \Large from Munich\\
    \vspace*{\stretch{2}}
    \Large Munich, #6
  \end{center}

  \newpage								
  \thispagestyle{empty}
	
	\textbf{Acknowledgements:}\vspace*{0.5cm}
	
	I wish to mention
	\begin{itemize}
		\item my advisor Prof. Ivo Sachs, to thank him for accepting my topic and making contact to Prof. Chamseddine, who will be the secondary judge at this thesis' defense.
		\item my family for providing me with support throughout my time making this thesis. In particular, I thank my father for many helpful on-topic conversations in which he probably learned more about physics than I did.
		\item my girlfriend Michaela for pulling me back on track during the many times I felt unfocused and for sharing her experience with me when I needed it.
		\item my friends Rex and Ikwik who corrected the entire, bloated script, even though this is not their area of expertise. Rex in particular read it multiple times, which I am highly thankful for.
		\item my many university colleagues for putting up with my field theoretic babble for three long years and pushing me to keep going when I did not want to, and also for keeping me down-to-earth at times. In particular, I thank Lukas and Rasmus for going over parts of the script and fixing some of my language.
	\end{itemize}

  \cleardoublepage
  \thispagestyle{empty}

  \vspace*{\stretch{1}}
  {\parindent0cm
  \rule{\linewidth}{.7ex}}
  \begin{center}
    \vspace*{\stretch{1}}
    \bfseries\Huge #1
    \vspace*{\stretch{1}}
  \end{center}
  \rule{\linewidth}{.7ex}

  \vspace*{\stretch{4}}
  \begin{center}
    \Large\textbf{Bachelorarbeit}\\
    \Large an der #4\\
    \Large der Ludwig--Maximilians--Universität\\
    \Large München\\
    \vspace*{\stretch{1}}
    \Large vorgelegt von\\
    \Large\textbf{#2}\\
    \Large aus #3\\
    \vspace*{\stretch{2}}
    \Large München, #6
  \end{center}

  \newpage
  \thispagestyle{empty}

  \vspace*{\stretch{2}}

  \begin{flushleft}
    \large Supervisor: #7 \\[1mm]
  \end{flushleft}

  \cleardoublepage
}

\begin{document}

\LMUTitle
      {Wechselwirkungen in MacDowell-Mansouri Gravitation}	
      {Simon Felix Langenscheidt}		
      {München}					
      {Fakultät für Physik}     
      {München 2019}            
      {\today}             		
      {Prof. Dr. Ivo Sachs}  

  \cleardoublepage
  \markboth{Abstract}{Abstract}
  	\vspace*{2cm}
	{\Huge \sffamily{\textbf{Abstract}}}\\
		
	In this bachelor thesis, possible kinetic terms and couplings of standard fields in MacDowell-Mansouri-Stelle-West gravity are studied with some aspects of group theory  in mind. Possible obstructions to these couplings are considered and used to make statements about the validity of the theory when coupled to matter. While interactions themselves turn out to be mostly unaffected except for scalar fields, the theory fails at its goal of putting gravity on equal footing with Yang-Mills theories. This happens with the kinetic term for spin-1 gauge fields and spin-0 ones, as one needs auxiliary fields to ensure manifest covariance with respect to the internal group $ SO(2,3) $.
	\vspace*{1cm}\\
	In dieser Bachelorarbeit werden mögliche kinetische Terme sowie Wechselwirkungen von Standardfeldern in MacDowell-Mansouri-Stelle-West Gravitation betrachtet. Hierbei werden gruppentheoretische Aspekte in Betracht gezogen. Es werden mögliche Einschränkungen auf die gefundenen Wechselwirkungen diskutiert und damit die Validität der Theorie eingeschätzt, wenn diese an Materie gekoppelt wird. Auch wenn Wechselwirkungen größtenteils unverändert bleiben, so müssen die kinetischen Terme von sowohl Spin-1 als auch Spin-0 durch Hilfsfelder augmentiert werden. Das Ziel des MacDowell-Mansouri-Formalismus, Gravitation und Yang-Mills Theorie anzugleichen, kann somit nicht erfüllt werden, wenn manifeste Kovarianz mit Bezug auf die interne $ SO(2,3) $ erreicht werden will.

  \renewcommand\contentsname{Table of contents}
  \tableofcontents
  \markboth{Table of contents}{Table of contents}

\mainmatter\setcounter{page}{1}

	\chapter{Preface}\label{Preface}
	After the undisputed success of general relativity (GR) or Einstein gravity (EG) at describing the intermediate- and large-scale behaviour of gravity, reformulations and extensions of said theories became paramount. While many formulations intend to modify the predictions of EG, others take a more conservative approach. Often, this happens in an attempt to find possible UV completions of gravity, for example by including additional terms in the equations of motion. In these formulations, one rewrites the theory to reach a form which gives the same predictions, but leads naturally to more generalised theories. The primary example of this approach is in the Palatini formulation of EG \cite{palatini}, which is equivalent to GR in regular cases, but canonically allows the coupling of fermionic matter to gravity. In this scenario, the presence of incompatible data (the existence of gravitating fermions and the absence of spinor representations of $ GL(n) $\cite{VasilievTheory}) leads to a different theory, which has other features, but can be related to the typical one under standard circumstances. This new formulation, however, has suggestive structures that did not exist in standard GR and make it more alike to gauge theories.\\
	Starting off there, more or less conservative extensions of EG exist, for example in the form of Einstein-Cartan theory\cite{CartanReview} or the Holst action\cite{holst}, which, in spirit, lead to even more elaborate reworks, extending even into quantal systems. In this way, a reformulation leads to new ideas by making use of different structures. From this perspective, MacDowell-Mansouri-Stelle-West (MMSW) gravity\cite{chamseddine,MM,West}, a reformulation of EG as a "\textit{quasi}-Yang-Mills theory", is worth considering due to its potential as a stepping stone to other, more interesting insights.\\
	However, most of the considerations of said theory are set in a pure gravity scenario. The issue of coupling it to matter has been raised before\cite{Wise} and recieved multiple answers\cite{Fukuyama}, most of which, though, often exclude an essential element of MMSW: Manifest gauge invariance under $ SO(2,3) $. In these coupling prescriptions, one either works purely in the reduced invariance phase, or one uses a reformulation of MMSW as a Background-Field theory with matter added as defects in the sense of condensed matter theory, which gives rise to well-known actions for point and string matter in a curved background.\cite{baez2007,Fairbairn_2008} As these methods demonstrate, a coupling is typically possible. However, in this thesis, I will look at this problem with a bit of representation theory in mind to elucidate some of the field content.
	The structure of this thesis is as follows: First, MMSW will be presented in the relevant form. Following on this, the representation theory of the internal invariance group will be laid out, then used to analyse the spin content of some fields. Then, individually, kinetic interaction terms will be considered for $ s= 0,\frac{1}{2} $ and $ s=1 $ gauge fields. As a conclusion, some problematic aspects of the framework are discussed.
	\chapter[MMSW Gravity]{MMSW Gravity}
	In this section, the setting of the investigations in this thesis is given. First, a quick review of the Palatini form of EG is given, after which MMSW will be presented independently in a metric-free form. Then, the relation between the two and GR will be shown and issues with this formalism will be deduced.
\section[Palatini gravity]{Prologue: The Palatini action}\label{Palatini}
	In the following, be $ dim(M) = n = 4 $ the dimension of spacetime.
	GR is a formalism of gravity which is based on the framework of Riemannian geometry. It centers on a pseudo-Riemannian metric that is defined on the tangent bundle $ TM $ of a spacetime manifold. From this metric, one generates a metric-compatible, torsion free, affine connection on $ TM $, the Levi-Civita connection, which gives notions of parallel transport of tensors of $ TM $.\cite{Nak} This viewpoint is successful due to its constrained nature and capability to give physical interpretations, but it lacks some features one might desire. In particular, there is no obvious way to include spinors of any kind in this scheme, as these do not transform as tensors of $ GL(n) $, and an affine connection does not allow for the transport of these objects.\cite{Nak,VasilievTheory}\\
	The Palatini formulation of EG changes this situation by ditching the metric tensor as a free variable. The intuition is as follows: One chooses an orthonormal frame at each point in spacetime, $ e_{a} = e_a^{\mu} \partial_{\mu} $, $ a \in \{0,1,2,3\} $, $ \mu $ a coordinate basis index, in which the metric tensor will have the component expression of flat spacetime, $ g(e_a,e_b) = \eta_{ab} $, at each point. One also introduces a similar set of one-forms $ \theta^a = \theta^a_{\mu} dx^{\mu}$ which form a so-called \textit{coframe}, which is required to satisfy $ \theta^a(e_a) = \delta^a_b $. With this choice, the cometric also satisfies $ g^{-1}(\theta^a,\theta^b) = \eta^{ab}$.\\
	The actual construction is slightly more geometrical. I follow the exposition of Wise\cite{Wise} and occasionally use identities from Nakahara.\cite{Nak} One introduces a real vector bundle of rank $ n $ called the \textit{fake tangent bundle }$\mathfrak{T}$, which is equipped with a fixed, canonical metric $ \eta $, the one of flat spacetime. Then, one introduces a vector bundle morphism $ \theta : TM \to \mathfrak{T} $, which one typically requires to be an isomorphism of vector bundles. When $ TM $ is trivialisable, $ \theta $ gives rise to a coframe, by restricting to a map $ \theta_p : T_pM \to \mathbb{R}^{1,3} \; \forall p \in M$. The metric on $\mathfrak{T}$ gives rise to one on $ TM $ via pullback through $ \theta $:
	\begin{equation}\label{2-1}
		g(u,v) := \eta(\theta(u),\theta(v))
	\end{equation}	or in components
	\begin{equation}\label{2-2}
		g_{\mu \nu} = \eta( \theta(\partial_{\mu} ),\theta(\partial_{\nu} ) ) = \eta_{ab} \theta^a_{\mu} \theta^b_{\nu}
	\end{equation}
	This expresses the metric in terms of the morphism $ \theta $. However, one still needs a notion of parallel transport which is metric-compatible. This is achieved, for example, by assuming the existence of a principal connection $ \omega $ on a principal $ SO(1,3) $-bundle over $ M $. By realising $\mathfrak{T}$ as a bundle associated to this principal bundle, one has a fitting, metric compatible connection on it, which may be pulled back  to $ TM $ just like $ \eta $. One thus has an $ \mathbb{R}^{1,3} $-valued one-form and an $ \mathfrak{so}(1,3) $-valued connection one-form on $ M $, which allow one to reconstruct the affine connection of GR, with the tetrad reconstructed from $ \theta $ and the connection coefficients from $ \theta $ and $ \omega $. Note that one does not a priori require the affine connection to be torsion-free. Indeed, the Palatini action will force the torsion to vanish when going on-shell in the vacuum. \\
	The connection to regular covariant differentiation is given by 
	\begin{equation}\label{2-2a}
		(\nabla_\mu V) = (\dot{D}_\mu V)^a e_a
	\end{equation}which uniquely connects the two connections via 
	\begin{equation}\label{2-2b}
		\Gamma^\rho_{\mu \sigma} = e^\rho_a (\dot{D}_\mu \theta)^a_\sigma = -\theta^\rho_a (\dot{D}_\mu e)^a_\sigma
	\end{equation}
	But one has to keep in mind that the two connections will only be equal on-shell.
	The Palatini action is
	\begin{equation}\label{2-3}
	S_{Pal} = \frac{1}{2\kappa} \int_{M}^{}  R\wedge \theta\wedge \theta = \frac{1}{2\kappa} \int_{M}^{} \varepsilon_{abcd} \theta^{a}\wedge \theta^{b}\wedge R^{cd}
	\end{equation}
	with $ R $ the curvature two-form of $ \omega $, given by $ R = d\omega + \frac{1}{2}{[\omega,\omega]} $. Also, the $ \wedge $ in (\ref{2-3}) acts as expected on the $ \mathbb{R}^{1,3} $ and $ so(1,3) $ Lie algebra parts of the forms involved.\\
	One may also include a cosmological constant term by 
	\begin{equation}\label{2-4}
	S_{Pal, \Lambda} = \frac{1}{2\kappa} \int_{M}^{} R\wedge \theta\wedge \theta - \frac{\Lambda}{6} \theta\wedge \theta\wedge\theta\wedge \theta
	\end{equation}
	This action has as its variables the bundle morphism $ \theta $, here seen as a $ \mathbb{R}^{1,3} $-valued one-form and an $ \mathfrak{so}(1,3) $-valued connection one-form, commonly known as the spin connection. The equations of motion give us 
	\begin{equation}\label{2-5}
		d_\omega(\theta\wedge\theta) = 0 
	\end{equation}
	\begin{equation}\label{2-6}
		\theta\wedge R -\frac{\Lambda}{3}\theta\wedge\theta\wedge\theta = 0
	\end{equation}
	where the former is achieved by variation in $ \omega $, the latter in $ \theta $. 
	In the case where $ \theta $ is an isomorphism, the first equation reduces to the vanishing of the torsion (of the associated affine connection on $ TM $). This is a requirement typical of GR. The second equation is just Einsteins field equation without sources. \\
	One can generically use this action instead of the usual Einstein-Hilbert action for EG by replacing the metric on $ M $ by appropriately set up copies of $ \theta $ and using the spin connection for covariant derivatives on $ TM $. In addition, one can couple spinorial fields to gravity with this by using $ \omega $ on spin bundles associated to the principal $ SO(1,3) $-bundle used in the construction, usually the frame bundle of $ M $. This is now possible as spin lifts of all required structures exist for the Lorentz group, while they did not for $ GL(4) $. However, one sees that once $ \omega $ is used for couplings in the rest of the action, there will generically be nonzero torsion, as the RHS of (\ref{2-5}) will not vanish. This is similar to the case of Einstein-Cartan theory\cite{TorsionDirac}, where torsion is algebraically related to the spin tensor of matter Lagrangians, while the canonical energy-momentum tensor takes the place of the Hilbert one. One can of course eliminate the torsion from the expressions by use of these relations, which return one to the standard GR expressions by reacquiring the Hilbert EM tensor via the Belinfante-Rosenfeld method\cite{MarsdenStressEn}. This will incorporate nonlinearities into the matter action, however. \\
	For example, for the Dirac Lagrangian, one has, instead of vanishing torsion:
	\begin{equation}\label{2-7}
		T^\mu_{ab} \propto \kappa \bar{\psi}\gamma^\mu \Sigma_{ab}\psi
	\end{equation}
	The details of this result will be covered later.\\
	This should give an overview of the Palatini form of EG. One should see immediately that the bundle morphism (or solder form) $ \theta $ takes on a very different role than the actual connection used in parallel transport. In fact, one might wonder if a description purely in terms of connections is possible. Obviously the spin connection will not suffice, as we had to use $ \theta $ in order to reconstruct the connection on $ TM $. However, the framework of Cartan connections allows one to at least make sense of $ \theta $ and eventually even take the step into a formulation with just Ehresmann connections.
		
\section[Formulation]{The MMSW formulation}
	Before I start presenting the modern formulation, a bit of historical context is in order. The deSitter groups were studied in their own rights in the 1920s and 30s, during which most of their properties were discovered.\cite{Dirac} Most of this interest was due to general activity in group representation theory in connection with the newly formulated Quantum mechanics, in which symmetry principles proved to be most easily studied using representations of groups. The main interest was in the deSitter isometry group $ SO(1,4) $ due to its similarity to the Lorentz group $ SO(1,3) $. The studies did not catch on and were laid to rest until the late 70s. During that time, supersymmetric gravity theories were created, but mostly by adding terms to Lagrangians iteratively until they achieved supersymmetry.
	This changed with a theory by Chamseddine\cite{chamseddine}, developed independently at the same time by MacDowell, Mansouri\cite{MM} and West\cite{West}, where the then simplest version of supergravity, containing only a gravitino and the graviton, was derived from a simple-looking action with internal group $ SP(4;\mathbb{R}) $ or its corresponding supergroup $ OSp(1,4) $. In Chamseddine's version, however, the action was not quadratic in the field strength as is typical of Yang-Mills theories. In this formulation, one could easily connect the two theories of GR and supergravity and also give regular EG the look of a gauge theory. This viewpoint was further investigated by Stelle and West\cite{Stelle}, which gave a more symmetry-based viewpoint and reinterpreted the necessary structures in terms of spontaneous symmetry breaking. This theory was later used as a base on which to study more recent ideas, such as dualities, massive gravity and topological gravity. For example, the supersymmetric version was later used to study Seiberg duality of gravity, and a dual was subsequently found\cite{SDual1,SDual2}. In all of these events, though, coupling to other fields was usually neglected, in particular in the more geometric studies. In recent years, the theory was fully connected to the framework in which it is to be understood, Cartan geometry\cite{Wise}.
\subsection[Cartan geometry]{Cartan geometry} \label{Cartan}
	The remarks of this section are loosely based on work of Wise \cite{Wise}, Nakahara\cite{Nak} as well as Stelle and West\cite{Stelle}. One might use Sharpe\cite{Sharpe} as a reference to Cartan geometry in general, but all necessary notions will be given here.\linebreak
	The basic idea of Cartan geometry is a generalisation of either Klein geometries to allow local curvature or of Riemannian geometry to allow the tangent spaces to be different from a flat space. Namely, we shall remember that a nonzero cosmological constant is equivalent to a constant background curvature of spacetime. As such, one might think that using Minkowski space as a tangent space at each point in the presence of a cosmological constant is not the optimal choice. Namely, it would be best if the geometry of spacetime were locally approximated by an (Anti-)deSitter space, which is a spacetime of constant curvature and a symmetric space. In the case that the real spacetime is an AdS space, one could identify all the tangent AdS spaces, just like how one can identify Minkowski space and its tangents. This idea is formalised by Cartan geometry, which allows one to replace tangent spaces by spaces of \textit{Klein geometries}. 
	These are pairs $ (G,H) $ of a Lie group and a closed subgroup of it, which can be used to form a principal $ H $-bundle $ G\stackrel{\pi}{\longrightarrow} G/H $ over the space of cosets of $ H $ in $ G $. One assumes in the typical construction that $ G/H $ is connected, which can always be done. 
	Now, the space $ G/H $ may be seen as a geometry which generalises the kind of trinity one has with flat Euclidean space, spheres and hyperbolic spaces; The group $ G $ will be the group of isometries of the space. For instance, the sphere may be realised as $ \mathbb{S}^n \cong O(n+1)/O(n) $, or hyperbolic space as $ \mathbb{H}^n \cong SO(1,n)/(O(1)\times O(n)) $ and most intuitively, Euclidean space as $ \mathbb{E}^n \cong (\mathbb{R}^n \rtimes O(n))/O(n)$. In the last case, one can see that effectively the rotations have been removed from the Euclidean group, leaving $ \text{SGal}(n) $ to look similar to $ \mathbb{R}^n $. In fact it is of the same dimension, but is the affine space of dimension n. In general, one can view $ H $ as the isotropy group of any point in the space $ G/H $. \linebreak
	In principle, this construction is independent of the groups $ G $ and $ H $, but we are interested in the special case where $ H = SO(1,3) $. This will ensure that our Klein space, which we use as a model space, will have at least the isotropy group of the regular tangent spaces. In addition, one will later gain a principal $ SO(1,3) $-bundle for other uses, which will be important in linking back to regular formulations.
	For the symmetry group $ G $ we have essentially three choices: The conformal group of flat spacetime, $ SO(2,4) $ and the two (Anti-)deSitter isometry groups $ SO(1,4) $ and $ SO(2,3) $, corresponding to positive and negative constant spacetime curvature each. One might consider other groups, but these are the only ones which naturally occur as symmetry groups of four-dimensional spaces and which are extensions to the Lorentz group. The choice of $ SO(2,4) $ will be a choice of conformal geometry as tangent spaces, and will not produce tangent spaces of the right dimension, as  $ dim(SO(2,4)/SO(1,3)) = 9 $, which can be deduced from simple counting of parameters. In addition, the conformal group can not be contracted to the Poincaré group via the usual process of group contraction. This reduces our choices to the (Anti-)deSitter isometry groups, of which we choose $ SO(2,3) $. This choice might at first seem arbitrary, but shall be explained later in terms of the representation theories of both groups. Namely, deSitter space does not admit positive energy representations, which is inadmissible for potentially quantisable theories.\cite{EvansIrreps} \newline
	Given these groups, we can define a Cartan geometry as follows:\newline
	A Cartan geometry (($ P $,$\, \pi \,$,$ M $),$\, \alpha $) modeled on $ G/H $ is a principal $ H $-bundle $ P\stackrel{\pi}{\longrightarrow} M $ with a $ \mathfrak{g} $-valued one-form called the \textit{Cartan connection form}, $\linebreak \alpha:TP\longrightarrow \mathfrak{g} $, which satisfies:
	\begin{enumerate}
		\item If $ V $ is a vertical vector field on $ P $, then $ \alpha(V) \in \mathfrak{h} $, where $ \mathfrak{h} $ is the Lie algebra of $ H $. In particular, on fibres of $ P $, $ \alpha $ restricts to the Maurer-Cartan form $ \omega_H $, which has, if $ \xi^{\#} $ is the vertical vector field generated by $ \xi \in \mathfrak{h} $: $ \omega_H(\xi^{\#} ) = \xi \;\;\;\forall \xi \in \mathfrak{h}$.\newline
		This means that $ \alpha $, apart from being $ \mathfrak{g} $-valued, is an Ehresmann connection on the principal bundle.\cite{Nak}
		
		\item $ (R_h)^\ast \alpha = Ad_{h^{-1}}\alpha $. In other words, $ \alpha $ transforms in the adjoint representation of the Lie group under right action/gauge transformations. If (1) meant $ \alpha $ was an Ehresmann connection, this makes it into a principal connection.\cite{Nak}
		
		\item $ \forall p \in P $, $ \alpha_p:T_pP \longrightarrow \mathfrak{g} $ is a vector space isomorphism. This condition may be relaxed to requiring the domain and codomain being of the same dimension, as shall be explained later. This is also known as the \textit{absolute teleparallelism} condition.
	\end{enumerate}
	
	The third condition is easily the most consequential one. It implies an isomorphism $ T_xM \cong \mathfrak{g}/\mathfrak{h}$, where the latter can be seen as the tangent space to $ G/H $. This gives rise to the intuitive picture that the spacetime $ M $ is modelled by $ G/H $, and thus has the same tangent spaces. In addition, the isomorphism allows for an injection $ X_\xi : \mathfrak{g} \longrightarrow \Gamma(TP)$, so that one can extend the $ \# $-map from the Lie algebra to vertical vector fields to more general vector fields on $ P $. Namely, the restriction to $ \mathfrak{h} $ gives the $ \# $-morphism, while the restriction to $ \mathfrak{p} := \mathfrak{g}/\mathfrak{h} $ gives vector fields "on $ M $". Indeed, the isomorphism can be inverted to give both vector fields on $ M $ or $ P $, the latter of which will be horizontal with respect to the $ M $-connection on $ P $, when we have the reductive case, defined as follows. \newline
	We talk of the \textit{reductive case} in a Cartan geometry when we have a decomposition $ \mathfrak{g} = \mathfrak{h} \oplus \mathfrak{p} $, which is invariant under action of $ Ad(H) $. This case is special as it allows for many structures on $ P $, for example a principal H-connection directly from the Cartan connection. 
	One central consequence of the reductive case is the splitting of $ \alpha $ into parts valued in $ \mathfrak{h} $ and $ \mathfrak{p} $, respectively. Namely,
	\begin{equation}\label{2-8}
		\alpha = \mathcal{\sigma} \oplus \frac{1}{\rho}\mathcal{\theta}
	\end{equation}
	with $ \mathcal{\sigma} : VP \longrightarrow \mathfrak{h} $ being a principal $ H $-connection on $ P $ due to the $ Ad(H) $-invariance of the splitting. It fulfils the extra condition of being an isomorphism of the vertical part of each tangent space of $ P $, $ V_pP \cong \mathfrak{h} $. This condition is essentially the same as requiring $ V_pP \cong H $, in accordance with P being a principal bundle.
	In addition, one can split the \textit{Cartan curvature} of $ \alpha $, defined analogously to the one for Ehresmann connections, into parts as well:
	\begin{equation}\label{2-9}
		\mathcal{F}= F[\alpha] := d\alpha +\frac{1}{2}{[\alpha,\alpha]} 
	\end{equation}
	\begin{equation}\label{2-10}
		= d\mathcal{\sigma} +\frac{1}{2}{[\sigma,\sigma]} + \frac{1}{\rho^2}\mathcal{\theta} \wedge \mathcal{\theta} \; \oplus \; \frac{1}{\rho}d \mathcal{\theta} +\frac{1}{\rho} {[\mathcal{\sigma},\mathcal{\theta}]} 
	\end{equation}
	\begin{equation}\label{2-11}
		= F[\mathcal{\sigma}] +\frac{1}{\rho^2}\theta\wedge\theta \, \oplus \, \frac{1}{\rho}d_\sigma\theta
	\end{equation}
	We will later see that $ F[\sigma] $ can be identified with the Riemann curvature of the tangent bundle, while $ T := d_\sigma\theta $ is the Torsion of the associated affine  connection on $ TM $. In the same way, the $ \theta\wedge\theta $-term will be a cosmological constant term, thus giving a concise way to integrate the three objects into one field strength of the same connection. One should note, of course, that $ \alpha $ is not an Ehresmann connection, which makes this field strength different in concept as well, but is precisely the reason it allows for this kind of decomposition. It should be noted that $ \mathcal{F} $ satisfies a decomposable Bianchi identity:
	\begin{equation}\label{2-12}
		d_\alpha \mathcal{F} = 0
	\end{equation} 
	\begin{align}\label{2-13}
	&d_\sigma F[\sigma] = 0  & d^2_\sigma \theta = F[\sigma]\wedge \theta&
	\end{align}
	which, when symbolically viewed as statements about an $ SO(1,3) $-connection and a coframe field of $ TM $, are just the standard Bianchi identities. \linebreak
	Until now, we do not have any structures except for vector fields on the tangent bundle of our base space determined by the Cartan connection $ \alpha $. If we wish to establish a metric tensor and a connection on it, we need additional structure on the complement of $ \mathfrak{h} $ in $ \mathfrak{g} $, $ \mathfrak{p} $. Say this subspace of $ \mathfrak{g} $ is endowed with an $ Ad(H) $-invariant symmetric bilinear form, which we suppose to be nondegenerate for our purposes. Then the isomorphism $ \alpha_p: T_xM \longrightarrow \mathfrak{p} $ allows us to pull back that form to any tangent space of $ M $. Similarly, the Killing form of $ \mathfrak{g} $ will give, as it is invariant, a similar form on $ T_pP $. If $ \mathfrak{g} $ is semisimple (as in the case of $ \mathfrak{so(2,3)} $), the Killing form is nondegenerate, and so one gets a bundle metric on $ \mathfrak{g} $ , $ \mathfrak{p} $ and $ T_xM $. It is instructive to see this for the case of this thesis, which I prepared myself. Be
	\begin{equation}\label{2-14}
		\mathcal{K}_{AB,CD} = (\eta_{AC}\eta_{BD}  - \eta_{AD}\eta_{BC})
	\end{equation}
	the Killing form in components as in the appendix, where $ A \in \{0,1,2,3,5\} $ and $ AB $ is antisymmetric. Then, \begin{equation}\label{2-15}
		g_p:= \, \mathcal{K} \circ (\alpha_p \otimes \alpha_p) : \, T_pP \otimes T_pP \longrightarrow \, \mathbb{R}
	\end{equation}
	is a bilinear, $ Ad(H) $-invariant second rank tensor on $ TP $. It is a metric as $ \mathfrak{so(2,3)} $ is semisimple. With a pullback of $ \alpha $ onto $ M $ through some section of $ P \longrightarrow M $, one can thus use $ g $ as a metric on $ TM $. Now consider the $ H $-connection induced splitting of vector fields on $ M $\cite{Nak}, 
	\begin{equation}\label{2-16}
		V \, = V^V \oplus V^H \text{ such that } V^V\text{a section of } VP, \; \sigma(V^H) = 0.
	\end{equation}
	Here and from now on, one has to see $ \sigma  ,\, \theta $ as pulled back forms on $ TM $, and the identities should be understood pointwise. If $ V^V = 0 $, so that $ V $ is a horizontal vector field, one has $ \alpha(V) = \sigma(V) \oplus \theta(V)  =  \theta(V)$.
	If $ V = V^{AB} e_{AB} $, with the latter a basis section of $ TP $, one may identify vertical vector fields with fields such that $ V^{5b} = 0$ and horizontal ones with fields satisfying $ V^{ab} = 0$, with $ a,b \in \{0,1,2,3\} $, by choosing an appropriate basis. This splitting is, in the reductive case, $ Ad(H) $-invariant, so it is independent of basis, but it is useful to stick to this form of the splitting as the reductive case becomes simpler with it.
	So one gets for the metric, properly normalised, on $ TM $, with $ V,W $ horizontal: 
	\begin{equation}\label{2-17}
		g(V,W) = \mathcal{K} \circ (\alpha(V) \otimes \alpha(W))
	\end{equation}
	or in a basis

	\begin{align}\label{2-18}
		\begin{split}
		g(V,W) &= (\eta_{AC}\eta_{BD} - \eta_{AD}\eta_{BC})\; \alpha^{AB}(V) \alpha^{CD}(W)\\
		&= (\eta_{AC}\eta_{BD} - \eta_{AD}\eta_{BC})\; \alpha^{5a}(V) \alpha^{5b}(W)\\
		&= (\eta_{55}\eta_{ab} - \eta_{5b}\eta_{a5})\; \alpha^{5a}(V) \alpha^{5b}(W)\\
		&= \eta_{55}\eta_{ab}\; \theta^a(V) \theta^b(W)\\
		&= \eta_{ab}\; \theta^a_\mu \theta^b_\nu V^\mu W^\nu 
		\end{split}
	\end{align}
	
	which is the well-known definition of the metric in terms of a local vierbein $ \theta^a_\mu $. More properly, using a section $ s: M \longrightarrow P $, we have 
	\begin{equation}\label{2-19}
		g(V,W) \, = \, \eta_{ab} \; s^{*}(\theta)^a_\mu \, s^{*}(\theta)^a_\nu \; V^\mu W^\nu
	\end{equation}
	which shows that the metric exists only locally in this way unless $ P $ is a trivial bundle.
	Since $ P $, with $ \mathcal{\sigma} $ as a principal connection, becomes a principal $ H $-bundle, this can be understood as a reduction of the structure group\cite{Reduction} of some other bundle to one with structure group $ SO(1,3) $, which we have with $ P $. This motivates the idea of achieving these structures from a bundle with structure group $ SO(2,3) $. In fact, as explained by Wise\cite{Wise}, one can extend $ \alpha $ to a principal $ G = SO(2,3)  $-bundle $ Q \stackrel{\pi}{\longrightarrow} M $ via the canonical injection $ \iota $ of $ H $-bundles when $ Q $ is viewed as an associated bundle to $ P $ with fibre $ G $. The resulting connection $ \omega $, is a \textit{principal $ G $-connection} that satisfies \begin{equation}\label{2-20}
	ker(\omega) \cap \iota_*(TP) = \emptyset
	\end{equation}
	where of course $ TP $ stands for any vector field on $ P $. Conversely, any principal $ G $-connection which satisfies (\ref{2-23}) will give rise to a Cartan connection on $ P $ with all the previously mentioned structures.\\
	 \\
	However concise and intuitive the Cartan geometric picture is, it is operationally unsatisfactory. To generate a sensible notion of parallel transport, one has to switch to the associated principal connection on $ Q $. As an associated problem, one cannot use the Cartan connection on associated bundles in an obvious way. Additionally, one can not think of lifts of the structure group of a Cartan connection, which one would certainly need to consider spin structures. As such, in the following we will mainly consider the principal $ G $-bundle $ Q $ and construct the associated Cartan geometry from it.\\
	 \\
	Now, consider the bundle with fibre $ G/H = AdS_4 $ associated to $ Q $ via $ Ad(G) $-action. This is the generalisation of the tangent bundle $ TM $ to Cartan geometry. Sections of this bundle assign a point in an Anti-deSitter space to each point in the base manifold, which may be seen as the point of tangency of the local copy of AdS with the manifold point. A parallel transport of a point $ u $ in this bundle along some path in $ M $ will act on the AdS-part of the point $ u $ as an $ SO(2,3) $-transformation, so it will be a Lorentz transformation of the local copies of AdS which leaves the points of tangency fixed, or it might move the point of tangency via $ SO(2,3) $-translations. In the case that the connection is flat and $ M = AdS_4 $, one can imagine a global section to just assign a point in AdS to the same one in its local copy. In this way, it can be identified with its tangent Anti-deSitter spaces.\\
	In general, the existence of a global section of this bundle $ Q \times_{Ad(H)} G/H  \cong Q/H$ is equivalent to the existence of a reduction of the structure group of $ Q $ to $ H = SO(1,3) $\cite{Reduction}, which itself is equivalent with the global existence of a metric tensor on $ TM $. As such, since one imposes the condition (\ref{2-20}) in addition to $ Q/H $ being trivial to gain a Cartan connection and thus a metric on $ TM $, one need only assume global sections of $ Q/H $ when we do the triviality of $ P $ and (\ref{2-20}). However, this question is of no real relevance to this thesis and I shall disregard the existence of global sections for the considerations to come.

	I will now give the definition of MMSW gravity.
	\newpage

\subsection{The action} \label{MMSW Action}
	This section shows my definition of MMSW, which is analogous to Wise's\cite{Wise}.
	In the following, be $ M $ the four-dimensional smooth spacetime manifold under consideration, $ Q \stackrel{\pi_Q}{\longrightarrow} M $ a principal $ G=SO(2,3) $-bundle with principal connection $ \omega $, locally expressed through $ \{\mathcal{A}_i\} $ over an open covering $ \{U_i\} $ via sections $ {s_i} $ of $ Q $, which will be suppressed for convenience. Also, be $ \tau:M \longrightarrow Q/H $ a section of the associated AdS-bundle, with $ H = SO(1,3) $, where the AdS spaces shall be realised as subsets of $ \mathbb{R}^5 $, so that the entire bundle is a subbundle of a trivial $ \mathbb{R}^5 $-bundle. Be the symbolic field strength of a Lie algebra-valued one-form $ F[\omega]:= d\omega + \frac{1}{2}{[\omega,\omega]} $, with $ d $ the appropriate exterior derivative. It is normalised as $ F = \frac{1}{2!2!}F^{AB}_{\mu\nu} \, M_{AB}\otimes dx^\mu\wedge dx^\nu $ due to antisymmetry in internal and external indices.\\
	Then the MMSW action is given by 
	
	\begin{align}\label{2-21}
		\begin{split}
		S_{MMSW} &= \frac{-1}{4\alpha \rho} \int_{M}^{} \text{tr}(F[\mathcal{A}]\wedge \circledast F[\mathcal{A}])\\
		&= \frac{1}{4\alpha \rho}\int_{M}^{} F[\mathcal{A}]^{AB}\wedge (\circledast F[\mathcal{A}])_{AB} \\
		&= \frac{1}{8\alpha \rho}\int_{M}^{} F[\mathcal{A}]^{AB}\wedge F[\mathcal{A}]^{CD} \tau^E \epsilon_{ABCDE} 
		\end{split}		
	\end{align}
	where the \textit{"internal Hodge dual"}, named in reference to Wise, is defined as
	\begin{equation}\label{2-22}
		(\circledast F)_{AB} := \frac{1}{(4-2)!}\epsilon_{ABCDE} F^{CD} \tau^E
	\end{equation} where $ \epsilon_{ABCDE} $ is the usual Levi-Civita symbol in 5D. Extension to forms with a different number of internal indices is analogous and more information can be found in the appendix. The relevant mass dimensions are: $ {[A]} = 0 = {[\theta_\mu]}, {[A_\mu]}=1 = {[\omega_\mu]}, {[F_{\mu\nu}]=2}  $. Note the following details:
	\begin{itemize}
		\item The absence of the metric determinant volume factor. I do not presuppose here that $ M $ is (pseudo-)Riemannian. 
		\item The field strengths here are two-forms on $ M $ and so generate a volume element on it, which provides the means of integration. They have the decomposition $ F = R + \frac{1}{\rho^2}\theta\wedge\theta \oplus \frac{1}{\rho} T$, where $ R $ is the $ SO(1,3) $ curvature which corresponds to the Riemann tensor and $ T $ the torsion.
		
		\item The presence of $ \tau $. This object was introduced by Stelle and West\cite{Stelle} and had been called \textit{the director}. It was used as a method of achieving spontaneous symmetry breaking, which required additional scalar terms in the action, implicitly necessitating the use of a metric. In the geometric view of Cartan geometry, $ \tau $ has a natural interpretation as a choice of tangency for the AdS-bundle, and so functions mostly as an auxiliary field with units $ {[\tau^A]} = -1 $.
		
		\item The indices of the field strengths and $ \tau $ are of the adjoint and the fundamental representation of $ G $, respectively. The action is invariant under right action of $ G $, or gauge transformations. As it is made from differential forms, it is also diffeomorphism invariant.
		
		\item The internal Hodge star replaces the usual Hodge dual of a two-form defined on a Riemannian manifold. While its meaning is not entirely clear, it is my interpretation that it is an induced dual from the 5D space in which the model AdS spacetimes are embedded, which would also explain the presence of the tangency map $ \tau $. See the appendix for more information.
		\item The dimensionless constant $ \alpha = \frac{\kappa}{\rho^2} $, with $ \kappa = \frac{8 \pi G_N}{c^4} $ the Einstein constant, $ \rho $ a constant with units of length, the radius of the internal AdS spaces. It is related to the cosmological constant by $ \Lambda = \frac{3}{\rho^2} $, so that $ \alpha = \frac{\Lambda\kappa}{3} $. If one takes the current value of the cosmological constant from $ \Lambda $CDM, this has a value of around $ 10^{-120} $. Since, in addition, there is a factor of $ \frac{1}{\rho} $ in front, this is like in the case of the Einstein-Hilbert or Palatini action, where one has a dimensionful prefactor indicating that this is an effective theory. However, as will be clear later, the factor of $ \frac{1}{\rho} $ will be in front of \textit{every} action if chosen appropriately, which gives the corresponding cutoff a special role. 
	\end{itemize}
	It is worth noting that in West and Stelle's work, the director $ \tau $ played a different role from the one presented here. Indeed, it was simply thought of as a mostly nondynamical field with values in $ \mathbb{R}^{2,3} $ with a potential of the form $ \lambda \, (\tau^A\tau_A - \rho^2 ) $. This potential would ensure that spontaneous symmetry breaking would occur in the model, giving rise to four nondynamical Goldstone fields in the process as deviations from the minimum occur. It is also a point of AdS there, but it not manifestly constrained to it. However, in this thesis we follow the novel interpretation that $ \tau $ is indeed a map into actual Anti-deSitter spaces, which make the constraint nondynamical in nature and do not allow or require an on-shell field interpretation. One thus has the constraint $ \tau^A\tau_A = \rho^2 $ everywhere. Finally, the original version had an additional term in the Lagrangian, which would require the use of a volume factor $ \sqrt{-g} $, of which none exist in MMSW gravity. Thus, while for convenience we represent the components of $ \tau(x) $ as a 5-vector, it really is a point in the embedded AdS space, requiring no constraints as it is a section of the appropriate bundle. Also, there was a construction by Randono\cite{Randono} who managed to achieve spontaneous symmetry breaking from $ SO(2,3) \text{ to }  SO(1,3)$ by introducing a fermion condensate as a natural method of achieving a $ \tau $-like object. Since I have a more geometric intuition for this object now, this perspective will not be of relevance here - though it is noteworthy that there are dynamical methods for this symmetry breaking. \\

	One can easily vary with respect to $ \mathcal{A}^{AB} $ to get the  equations of motion. I find that they are
	\begin{align}\label{2-22a}
		\frac{2}{4\alpha \rho} (D\circledast \mathcal{F})_{AB} =0
	\end{align}
	
	where, in the \textit{standard gauge} $ \tau^A = \rho \delta^A_5 $,  which represents a phase of broken invariance or just the \textit{broken phase}, one has
	\begin{align}\label{2-23}
		(D\circledast \mathcal{F})_{AB} &= d(\circledast \mathcal{F})_{AB} + f_{AB,CD}^{EF} \mathcal{A}^{CD}\wedge(\circledast \mathcal{F})_{EF}\\
		(D\circledast \mathcal{F})_{5a} &= \frac{1}{\rho}\theta^b\wedge(\circledast \mathcal{F})_{ab} = \frac{1}{2}\epsilon_{abcd} \, \theta^b\wedge \mathcal{F}^{cd}\\
		(D\circledast \mathcal{F})_{ab} &= (\dot{D}\circledast \mathcal{F})_{ab} =d(\circledast \mathcal{F})_{ab} + f_{ab,cd}^{ef} \mathcal{A}^{cd}\wedge(\circledast \mathcal{F})_{ef}
	\end{align}as well as 
	\begin{equation}\label{2-23a}
	(\circledast \mathcal{F})_{ab} = \rho \epsilon_{abcd}R^{cd} +\frac{1}{\rho} \epsilon_{abcd} \theta^c\wedge \theta^d
	\end{equation}Also the Bianchi identity
	\begin{equation}\label{2-24}
		\dot{D}R = 0 \Leftrightarrow dR^{ab} + f_{cd,ef}^{ab} \omega^{cd}\wedge R^{ef} = 0
	\end{equation}holds.	
	By applying the internal Hodge star to the EoMs, one finds 
	\begin{equation}\label{2-25}
		\frac{1}{2\alpha \rho} (\circledast D\circledast \mathcal{F})^{AB} = \frac{\rho}{\alpha} (\circledast^2d\mathcal{F})^{AB} = 0
	\end{equation}which, in the standard gauge a using a formula in the appendix shows that
	\begin{align}\label{2-26}
	&\qquad d\mathcal{F}^{ab} = 0 \qquad&\Leftrightarrow \qquad  dR^{ab} = \frac{-1}{\rho^2} d(\theta^a\wedge \theta^b)&\\
	\end{align}
	One has to keep in mind here that applying the internal Hodge star removes the $ 5a $-part in the standard gauge so that, while the above equations certainly are correct, they are not equivalent to the full EoMs. Instead, the equivalent set is 
	\begin{align}\label{2-27}
	& \frac{\rho}{\alpha}d\mathcal{F}^{ab} = 0 & \epsilon_{abcd} \, \theta^b\wedge \mathcal{F}^{cd}=0 &	
	\end{align}
	Quite surprisingly, one sees that the vacuum equations of gravity are given by an exactness condition. If one has sources that only depend on the vierbein part $ \theta $, but not the spin connection part, one will still have the first equation, while the second will admit nontrivial solutions. In that case, one will have $ \mathcal{F} \in H^2(M;\mathfrak{so}(2,3))$.
	In vacuum, however, one should see that $ \mathcal{F} \equiv 0$ is the only solution. This, however, does not make the $ SO(1,3) $-part vanish; Instead one has 
	\begin{align}\label{2-28}
		R^{ab}= \frac{-1}{\rho^2}\theta^a\wedge\theta^b 
	\end{align}
	which is precisely the Riemann curvature from the EG solution which is Anti-deSitter space with radius of curvature $ \rho $. As such, the action is constructed such that AdS will give a finite action, as well as showing that $ \mathcal{F} $ encodes the deviation from Anti-deSitter curvature. These solutions can, for example, give rise to the potential given in Poincaré coordinates $ (z,x^i) $\cite[][(see page $ 29 $ of ref)]{VasilievTheory}
	\begin{align}\label{2-29}
	&\theta^a = \frac{1}{z}\delta^a_\mu dx^\mu & \omega^{ab} = -\frac{1}{\rho} (\delta^a_\mu \eta^{bz} - \delta^b_\mu \eta^{az}) dx^\mu&
	\end{align}
	It is clear that in the standard gauge, the torsional part of the field strength does not contribute to the action, as the internal Hodge star removes its contributions. Since the gauge can be chosen arbitrarily, as is seen in the next section, this is always the case. This is in constrast to the Palatini version, where torsion could be nonzero. This highlights a pathology in MMSW that will reappear later during the study of torsion. For now, one can find the torsion by expanding the full invariant action into its $ R,\theta, T $ subparts and then applying the variation with the Bianchi identity in mind. One finds that
	\begin{align}\label{2-30}
		\frac{1}{8\alpha\rho} \left( \theta_b\wedge \epsilon_{arst} \mathcal{F}^{rs}\tau^t  + 4 \dot{D}(\tilde{T})_{ab}\right) + \frac{1}{2\kappa \rho}\epsilon_{abcd}\theta^c\wedge T^d \tau^5+ \frac{\delta S_M}{\delta\omega^{ab}} = 0
	\end{align}where $ \tilde{T}_{ab} := \epsilon_{abcd} \tau^c T^d $. In the standard gauge, this reduces to simply
	\begin{align}\label{2-31}
		\epsilon_{abcd}\theta^c\wedge T^d + 2\kappa\frac{\delta S_M}{\delta\omega^{ab}} = 0
	\end{align} so that one can see clearly how the torsion has to vanish in vacuum and even for spinless matter. Still, it is not clear how this is compatible with the exactness condition from above. It seems that one needs to remove the part of (\ref{2-23}) quadratic in the Riemann curvature to get the condition on the torsion, even though it does not appear in the gauge fixed action in the first place.
	
\subsection[Connection to Einstein gravity]{Connection to Einstein gravity}\label{Limit}
	Here, the procedure to obtain the Palatini action from MMSW will be outlined. In this process, the power of this formulation will become apparent, as it directly gives, with no additional free parameters, a prediction that the Euler density will be important for quantisation while giving the exact couplings in question from $ \alpha $ and $ \rho $. It also neatly generates a cosmological constant term.\\
	To start off, one has to investigate $ \tau $. It is a "gauge" degree of freedom in MMSW in the sense that it does not affect the dynamics. The choice is one of local sections of the principal $ SO(2,3) $-bundle, which are used to pull back the forms from said bundle to $ M $. One may change said sections by a right action on the principal bundle, which is described on $ M $ by a gauge transformation\cite{Nak}:
	\begin{equation}\label{2-32}
		\mathcal{A} \mapsto g^{-1}(\mathcal{A} + d)g = g^{-1}\mathcal{A}g + g^{-1}dg
	\end{equation} where $ g $ is an $ SO(2,3) $-valued function on $ M $, used to change the section. Within the matrix representations, this is expressed as
	\begin{equation}\label{2-33}
	\mathcal{A}^A_B \mapsto (\Lambda^{-1})^A_{\;\, C}(\mathcal{A}^C_{\;\, D} + \delta^C_{\;\, D} d)\Lambda^D_{\;\, B} = (\Lambda^{-1})^A_{\;\, C}\mathcal{A}^C_{\;\, D}\Lambda^D_{\;\, B} + (\Lambda^{-1})^A_{\;\, C}d\Lambda^C_{\;\, B}
	\end{equation}
	with $ \Lambda $ a transformation matrix. This is for the up-down placement of indices in the adjoint representation; Here, the all-up version is conventionally used, which can be found from this using the defining property of matrices of $ SO(2,3) $.
	\begin{equation}\label{2-34}
	\mathcal{A}^{AB} \mapsto (\Lambda^{-1})^A_{\;\, C} \mathcal{A}^{CD} (\Lambda^{-1})^B_{\;\, D} + (\Lambda^{-1})^A_{\;\, C} d\Lambda^{CB}
	\end{equation}
	In addition, $ \tau $ transforms in the fundamental representation:  \begin{align}\label{2-35}
	\tau_A &\mapsto (\Lambda^{-1})^A_C \tau^C  \text{and F in the adjoint:}\\
	\mathcal{F}^{AB} &\mapsto (\Lambda^{-1})^A_{\;\, C} \mathcal{F}^{CD} (\Lambda^{-1})^B_{\;\, D}
	\end{align}
	Thus, the MMSW action is invariant under changes of local sections:
	\begin{align}\label{2-36}
		&S = \frac{1}{8\alpha\rho} \int_{M}^{} \mathcal{F}^{AB}\wedge\mathcal{F}^{CD}\tau^E \epsilon_{ABCDE} \\
		&\mapsto \frac{1}{8\alpha\rho} \int_{M}^{} \mathcal{F}^{KL}\wedge\mathcal{F}^{MN}\tau^O \epsilon_{ABCDE} (\Lambda^{-1})^A_{\;\, K} (\Lambda^{-1})^B_{\;\, L} (\Lambda^{-1})^C_{\;\, M} (\Lambda^{-1})^D_{\;\, N} (\Lambda^{-1})^E_{\;\, O}\\
		&=\frac{1}{8\alpha\rho} \int_{M}^{} \mathcal{F}^{KL}\wedge\mathcal{F}^{MN}\tau^O \epsilon_{KLMNO} \text{det}(\Lambda^{-1}) \; = \; S
	\end{align}
	as all elements of $ SO(2,3) $ have determinant $ 1 $.\\
	The choice of $ \tau $ is in principle independent of the local section chosen, but expressed with respect to it, similar to the situation of the other variables. Using this gauge invariance, one has the freedom to make any given section $ \tau $ assume a standard form by an appropriate local transformation. It is this standard form which leads one back to the Palatini action. Choose the gauge so that $ \tau^A(x) \equiv  \rho \delta^A_5  $. Then, one has
	\begin{align}\label{2-37}
		S &= \frac{1 }{8\alpha} \int_{M}^{} \mathcal{F}^{AB}\wedge\mathcal{F}^{CD} \epsilon_{ABCD5}\\
		&= \frac{1}{8\alpha} \int_{M}^{} \mathcal{F}^{ab}\wedge\mathcal{F}^{cd} \epsilon_{abcd}
	\end{align}
	 where $ a,b,c,d \in \{0,1,2,3\} $, which are indices of the Lorentz algebra.
	 One can already see here that through this choice of gauge, the torsion does not contribute at all to the action. However, the degrees of freedom associated to it did not disappear. Once one makes a transformation to another gauge in which the torsion contributes, the field strength will change accordingly to compensate for this change. For example, by changing to the $ \tau^A(x) \equiv \rho \delta^A_0 $-gauge, one eliminates the $ \mathcal{F}^{0b} $-part of the field strength, while the torsion will absorb exactly those components under the gauge transformation.\\
	 Now one has to use the decomposition into $ \mathfrak{h} $ and $ \mathfrak{p} $ parts which can be interpreted through the Cartan geometry arising from the principal connection $ \omega $ on $ Q $. Suppose one has a principal $ SO(1,3) $ bundle $ P $ over $ M $ which is a reduction of $ Q $. This may be achieved by the existence of a global section of the AdS-bundle. Then one can pull back $ \omega $ to $ P $, giving a Cartan connection on it, provided the constraint (\ref{2-20}) is satisfied. This gives all the structure one needs to construct a metric connection on $ TM $ from $ \omega $. One has for the curvature the decomposition:
	\begin{equation}\label{2-38}
		\mathcal{F}^{ab} = \mathcal{R}^{ab} + \frac{1}{\rho^2} \; \theta^a\wedge\theta^b
	\end{equation}
	where $ \theta^a_\mu $ is unitless and will be identified with the vierbein. Using this in the action, one finds
	\begin{align}\label{2-39}
		\begin{split}
		S &= \frac{1 }{8\alpha} \int_{M}^{} (\mathcal{R}^{ab} + \frac{1}{\rho^2} \; \theta^a\wedge\theta^b)\wedge(\mathcal{R}^{cd} + \frac{1}{\rho^2} \; \theta^c\wedge\theta^d) \epsilon_{abcd}\\
		&= \frac{1 }{8\alpha} \int_{M}^{} (\mathcal{R}^{ab}\wedge \mathcal{R}^{cd} + \mathcal{R}^{ab}\wedge\frac{1}{\rho^2} \; \theta^c\wedge\theta^d +\\
		& \qquad \frac{1}{\rho^2} \; \theta^a\wedge\theta^b\wedge\mathcal{R}^{cd} + \frac{1}{\rho^4} \; \theta^a\wedge\theta^b\wedge \; \theta^c\wedge\theta^d) \epsilon_{abcd}\\
		&= \frac{1 }{8\alpha} \int_{M} \mathcal{R}^{ab}\wedge \mathcal{R}^{cd} \epsilon_{abcd}  \\
		&+ \frac{2 }{8\alpha\rho^2} \int_{M} \mathcal{R}^{ab}\wedge \; \theta^c\wedge\theta^d \epsilon_{abcd} + \frac{1}{2\rho^2} \; \theta^a\wedge\theta^b\wedge \; \theta^c\wedge\theta^d \epsilon_{abcd}
		\end{split}
	\end{align}
	Consider the terms in the second integral now. The second term is
	\begin{align}\label{2-40}
		\theta^a\wedge\theta^b\wedge \; \theta^c\wedge\theta^d \epsilon_{abcd} = \theta^a_{\mu}\theta^b_{\nu}\theta^c_{\kappa}\theta^d_{\lambda} \epsilon_{abcd} dx^{\mu}\wedge dx^{\nu}\wedge dx^{\kappa}\wedge dx^{\lambda}
	\end{align}
	Using that $ M^a_uM^b_vM^c_wM^d_x \epsilon_{abcd} = \epsilon_{uvwx} \text{det}(M)$, this results in
	\begin{align}\label{2-41}
		\text{det}(\theta)\epsilon_{\mu\nu\kappa\lambda}dx^{\mu}\wedge dx^{\nu}\wedge dx^{\kappa}\wedge dx^{\lambda} = 4! \text{det}(\theta) d^4x = 4! \, vol
	\end{align}
	with the determinant understood as the determinant of the matrix with entries $ \theta^a_\mu $. The first term, however, is 
	\begin{align}\label{2-42}
		\mathcal{R}^{ab}\wedge \theta^c\wedge\theta^d \epsilon_{abcd} = \mathcal{R}^{uv}\wedge \delta^a_u\delta^b_v\theta^c\wedge\theta^d  \epsilon_{abcd}
	\end{align}
	Suppose now that $ \theta^a_\mu $ is invertible, which is true in all but a degenerate case. This is also guaranteed when the principal connection on $ Q $ gives rise to a Cartan connection (which is an isomorphism of $ T_pP $ and $ \mathfrak{g} $). We then use the vierbein vector field $ e_a^\mu $, which is the dual of $ \theta^a_\mu $, to recover the curvature scalar: 
	\begin{align}\label{2-43}
		\begin{split}
		&\mathcal{R}^{uv} \wedge \delta^a_u\delta^b_v \theta^c\wedge\theta^d  \epsilon_{abcd}\\ 
		&= \mathcal{R}^{uv}_{\;\;\mu \nu} e^\rho_u e^\sigma_v\theta^a_\rho\theta^b_\sigma\theta^c_\kappa\theta^d_\lambda dx^{\mu}\wedge dx^{\nu}\wedge dx^{\kappa}\wedge dx^{\lambda} \epsilon_{abcd}\\
		&= \frac{1}{2}\mathcal{R}^{uv}_{\;\;\mu \nu} e^\rho_u e^\sigma_v dx^{\mu}\wedge dx^{\nu}\wedge dx^{\kappa}\wedge dx^{\lambda} \epsilon_{\rho\sigma\kappa\lambda} \text{det}(\theta)\\
		&= \frac{1}{2}\mathcal{R}^{\rho \sigma}_{\;\;\mu \nu}\epsilon^{\mu\nu\kappa\lambda} \epsilon_{\rho\sigma\kappa\lambda} \text{det}(\theta) d^4x\\
		&= \frac{2! \, 2!}{2} \, \mathcal{R}^{\rho \sigma}_{\;\;\mu \nu} \, (\delta^\mu_\rho \delta^\nu_\sigma -\delta^\mu_\sigma \delta^\nu_\rho) \text{det}(\theta) d^4x\\
		&=  2! \, \mathcal{R}^{\rho \sigma}_{\;\;\rho \sigma} \,\text{det}(\theta) d^4x
		= 2 R  \,\text{det}(\theta) d^4x = 2 R \, vol
		\end{split}
	\end{align}
	once one remembers that $ \text{det}(\theta) = \sqrt{-\text{det}(g)} $ and our choice of $ \alpha = \frac{\kappa}{ \rho^2 } $, we have the Einstein-Hilbert action.
	\begin{align}\label{2-44}
		\begin{split}
		S_{MMSW} &= \frac{1 }{8\alpha} \int_{M} \mathcal{R}^{ab}\wedge \mathcal{R}^{cd} \epsilon_{abcd}  
		+ \frac{2}{8\alpha\rho^2} \int_{M} d^4x \sqrt{-g}( 2 R  + \frac{4!}{2\rho^2})\\
		&= \frac{1 }{8\alpha}\int_{M} R\wedge R  
		+ \frac{ 1 }{2\alpha\rho^2} \int_{M} d^4x \sqrt{-g}(  R  + 2\frac{3}{\rho^2})\\
		&= \frac{1 }{8\alpha} \int_{M} R\wedge R  
		+ \frac{1}{2\kappa} \int_{M} vol \,(  R  - 2\Lambda)
		\end{split}
	\end{align}
	where the identification $ \Lambda = \frac{-3}{\rho^2} $ has been made. 
	The power of the MMSW formalism can be seen here. It is equivalent to Einstein-Hilbert on the level of the action up to a term quadratic in the $ SO(1,3) $-curvature, which may be identified as the Euler class\cite{Nak} of the $ SO(1,3) $-bundle $ P $ and integrates to $ \frac{4\pi^2}{\alpha}\chi(M) $, where $ \chi(M) $ is the Euler characteristic. It is a topological term, which makes the MMSW formalism classically equivalent to Einstein-Hilbert or Palatini. MMSW thus predicts a relation between the coupling to a generic curvature squared term and the gravitational constants. This might in principle lead to a way to falsify MMSW as a model, by finding experimental values for these couplings and comparing them to theory. Namely, MMSW predicts $ c_{R^2} \approx 8\times10^{120} $. 
	\\
	It is interesting to note that MMSW bears similarities to the 5-dimensional Lovelocke gravity term:\cite{WZWMM}
	\begin{align}\label{2-45}
		\textit{vol}\, \mathcal{L}_{LL,5D} = \epsilon_{ABCDE}(\alpha_0\theta^A\theta^B\theta^C\theta^D\theta^E + \alpha_1  \theta^A \theta^B \mathcal{R}^{CD} \theta^E + \alpha_2\theta^A \mathcal{R}^{BC}\mathcal{R}^{DE})
	\end{align}	
	where the wedge product is suppressed. If one were to replace $ \theta^a $ by $ \tau^A $, and fix the constants appropriately, this would just be the MMSW action integrand. This, by itself, is also similar to the \textit{Chern-Simons} 5-form
	\begin{align}\label{2-46}
		\text{tr}(\mathcal{R}\wedge\mathcal{R}\wedge\theta -\frac{1}{2}\mathcal{R}\wedge\theta\wedge\theta\wedge\theta +\frac{1}{10}\theta\wedge\theta\wedge\theta\wedge\theta\wedge\theta)
	\end{align}
	This is a topological theory in 4+1D, suggesting that gravity in 3+1D might be mostly topological in nature; A hypothesis supported by the deformed BF theory reformulation of MMSW proposed by Freidel and Starodubtsev\cite{BFFreidel}. Additionally, as has been demonstrated by Anabalón\cite{WZWMM}, one can derive MMSW for $ ISO(1,4) $ from a nonlinear realisation of a scalar field (more precisely, a 0-form) valued in the adjoint representation and an $ ISO(1,4) $-connection like the one used here, in an action he calls a gauged Wess-Zumino-Witten model. There, one can impose a restriction on the 0-form $ h = \text{exp}(\phi) = \text{exp}(\frac{\phi^{AB}}{2}M_{AB}) $ by setting $ \frac{1}{2}\phi^{AB}\phi_{AB} = m^2 $, which preserves the gauged invariance. One can relate this to the current scenario by requiring that $ h = \text{exp}(\tau^A P_A) $, with $ P_A $ is some subset of the generators of $ ISO(1,4) $ with $ P_\mu $ proportional to the translation subalgebra. Then, the constraint becomes $ \tau^A\tau_A = m^2$, which is the one we seek. In a way, one could see MMSW as having a hidden, even larger symmetry under $ ISO(2,3) $ or perhaps even $ SO(2,4) $, but realised in a more complicated model. I will not go into more detail about this here.
	However, it is also noteworthy that a generalisation of CS forms  to any dimension admits a direct correspondence to MMSW. These Higgs-Chern-Simons-forms \cite{GeneralCSForms}(in particular see ($ 13 $) and ($ 34 $) of this reference) give more meaning to $ \tau $ and further cement its connection to 5D Chern-Simons forms and topological theories. More concretely, it seems that MMSW is a \textit{Higgs-Chern-Pontrjagin}-density derived from the 6D Chern-Pontrjagin class, with the only difference being the dimensions of $ \tau $. In this context, the choice of gauge group $ SO(p,5-p) $ becomes more natural.

	Now recall the equations of motion in the standard gauge:

	\begin{align}\label{2-47}
		& (d\mathcal{F})^{ab}+ \frac{\kappa}{\rho^2}\epsilon^{abcd} \frac{\delta S_M}{\delta \omega^{cd}}= 0 &   \epsilon_{abcd}\theta^b\wedge F^{cd} +2\kappa \frac{\delta S_M}{\delta \theta^a}=0&
	\end{align}
	The second equation will yield the Einstein field equations. Taking a spacetime component of it and dropping the forms, it reads
	\begin{align}\label{2-48}
		\epsilon_{abcd}\epsilon^{\mu\nu\kappa\lambda}\theta^b_\nu F^{cd}_{\kappa \lambda} +2\kappa \frac{\delta S_M}{\delta \theta^a_\mu}=0
	\end{align}
	One multiplies and sums both sides by $ \theta^k_\mu $ and converts some indices to find 
	\begin{align}\label{2-49}
		\epsilon_{abcd}\epsilon^{\mu\nu\kappa\lambda}\theta^k_\mu\theta^b_\nu \theta^r_\kappa \theta^s_\lambda F^{cd}_{rs} +2\kappa \frac{\delta S_M}{\delta \theta^a_\mu}\theta^k_\mu &=0\\
		\epsilon_{abcd}\epsilon^{kbrs}\det(\theta) F^{cd}_{rs} +2\kappa \frac{\delta S_M}{\delta \theta^a_\mu}\theta^k_\mu &=0
	\end{align} where the determinant identity $ \epsilon^{\mu\nu\kappa\lambda}\theta^k_\mu\theta^b_\nu \theta^r_\kappa \theta^s_\lambda = \epsilon^{kbrs}\det(\theta) $ was used. Using yet another Levi-Civita identity from the appendix one finds 
	\begin{align}\label{2-50}
		\det(\theta) (2F^{cd}_{\;\;\;cd}\delta^k_a + 2F^{kd}_{\;\;\;da} +2F^{ck}_{\;\;\;ac}) +2\kappa \frac{\delta S_M}{\delta \theta^a_\mu}\theta^k_\mu &=0\\
		\det(\theta) (F^{cd}_{\;\;\;cd}\delta^k_a -2F^{ck}_{\;\;\;ca}) +\kappa \frac{\delta S_M}{\delta \theta^a_\mu}\theta^k_\mu &=0
	\end{align}

	At this point one introduces an energy-stress tensor, but it is not the Hilbert one used in the EFEs. Instead, it will be related to the Hilbert one via the Belinfante-Rosenfeld procedure\cite{MarsdenStressEn}. Be it defined by $ T^{(0)a}_b \theta^b_\mu := \frac{-2}{\text{det}(\theta)} \frac{\delta S_M}{\delta e_a^\mu} $. By local Lorentz invariance, this needs to be symmetric in its indices. We can relate this definition to the RHS by employing the chain rule and the defining property of the vierbein $ \theta^a_\mu e^\beta_a = \delta^\beta_\mu$:
	\begin{align}\label{2-51}
		0 &= \frac{\delta (\delta^\beta_\mu)}{\delta \theta^{b}_\rho} = \frac{\delta (\theta^a_\mu e^\beta_a)}{\delta \theta^{b}_\rho}
		=\theta^a_\mu\frac{\delta ( e^\beta_a)}{\delta \theta^{b}_\rho} + \frac{\delta (\theta^a_\mu )}{\delta \theta^{b}_\rho}e^\beta_a\\
		&=\theta^a_\mu\frac{\delta ( e^\beta_a)}{\delta \theta^{b}_\rho} + \delta^\rho_\mu e^\beta_b \Rightarrow \frac{\delta ( e^\beta_a)}{\delta \theta^{b}_\rho} = - e^\beta_b e^\rho_a
	\end{align}
	and so
	\begin{align}\label{2-52}
		\frac{\delta S_M}{\delta \theta^{b}_\rho}  &= \frac{\delta ( e^\beta_a)}{\delta \theta^{b}_\rho} \frac{\delta S_M}{\delta e^\beta_a} = 2e^\beta_b e^\rho_a \; \text{det}(\theta) \; T^{(0)a}_c \theta^c_\beta =  2  \text{det}(\theta) \; T^{(0)a}_b \; e^\rho_a \\
	\end{align}
	So now the equation of motion can be written as 
	\begin{align}\label{2-53}
		(F^{cd}_{\;\;\;cd}\delta^k_a -2F^{ck}_{\;\;\;ca}) +2 \kappa  \; T^{(0)k}_a &=0\\
		F^{ck}_{\;\;\;ca} -\frac{1}{2}F^{cd}_{\;\;\;cd}\delta^k_a =  \kappa  \; T^{(0)k}_a 
	\end{align}
	And this already suggestive expression becomes 
	\begin{align}\label{2-54}
		\text{Ric}^k_a +\frac{3}{\rho^2}\delta^k_a -\frac{1}{2}R\delta^k_a -\frac{12}{2\rho^2}\delta^k_a &=  \kappa  \; T^{(0)k}_a\\
		G^k_a -\frac{3}{\rho^2}\delta^k_a =  \kappa  \; T^{(0)k}_a
	\end{align} by use of $ F^{ab}_{\;\;\;rs} = R^{ab}_{\;\;\;rs} +\frac{1}{\rho^2} (\delta^a_r\delta^b_s - \delta^a_s\delta^b_r) $. This is just Einstein's field equation in the mixed form, with cosmological constant $ \Lambda = -\frac{3}{\rho^2} $. 
	For matter that does not couple to the spin connection $ \omega $, this is all one needs.\\
	In that case, the vacuum equations (when viewed in the Palatini form) clearly imply that the torsion vanishes on-shell. Thus, the spin connection is uniquely fixed by the vierbein and one completely recovers the coupling description of GR. However, usually one does have dependence on the spin connection, as for example in actions of fields with spin. Then, the torsion will become nonzero according to (\ref{2-31}). This issue will be discussed in detail in section \ref{Torsion}.
	Note that this also induces nonlinearities in the matter fields, which will give rise to new dynamics. In the non-effective situation, however, these are absent.
	
	Thus, one goes through the following steps to reduce MMSW to Einstein-Cartan gravity:
	\begin{enumerate}\label{2-55}
		\item One starts with an $ SO(2,3) $-connection on $ Q $ and chooses a section $ \tau $ that reduces the bundle $ Q $ to $ P $, which is a principal $ SO(1,3) $-bundle.
		\item One chooses the standard gauge, which projects the action onto its $ SO(1,3) $-parts.
		\item One pulls back the connections on $ Q $ back to $ P $, which gives rise to a Cartan geometry modeled on $ AdS_4 $. 
		\item Then, one constructs a fake tangent bundle $ P \times_{SO(1,3)} \mathfrak{g}/\mathfrak{h} $ and uses sections of $ P $ to pull back $ \theta \text{ and } \sigma $ to a solder form and a one-form on $ TM $, respectively. 
		\item One also inherits a metric $ g $ on the fake tangent bundle from the Cartan geometry, giving rise to a metric on $ TM $.
		\item One algebraically eliminates the torsion through the field equations and thus relates $ \sigma $ to $ \theta $ and thus $ g $.
		\item An affine connection can then be constructed from $ \sigma $ and $ \theta $, which can be used to define parallel transport on sections of $ TM$.
		\item Variation with respect to $ g $ then gives the Einstein-Cartan field equations.
	\end{enumerate}
	This section is now closed with a few words on the issues of this theory.

\section[Issues]{Issues with the theory} \label{Issues}
	The most obvious problem with this setup should be the sign of the cosmological constant. The current acceleration rate of the scale factor of the observable universe suggests that we live in deSitter space. So a most natural proposition, given the experimental data, would be to consider $ SO(1,4) $ instead. Still, one might consider, from a theoretical standpoint, the positive energy representations and natural nonzero masses and connection to the AdS-CFT correspondence as virtues. These might warrant trying to fit a negative cosmological constant to data.
	In that case, the idea is to arrive at an effective positive cosmological constant from the base negative cosmological constant and a positive vacuum energy due other fields, e.g scalar fields, or perhaps the constant terms arising from canonical quantisation of other fields. This will, of course, result in a horrific fine-tuning problem, which is already well-known as the vacuum catastrophe. However, in the light that a negative cosmological constant might exist, it might seem that, given a better understanding of vacuum energies, one might be able to arrive at a natural way of almost cancelling the effective vacuum energy. This, in all generality, is outside of the scope of this paper. However, Fukuyama \cite{Fukuyama} has found a model in which conformal symmetry is broken by a Higgs-like mechanism and exactly cancels the background cosmological constant by its vacuum expectation value. The real cosmological constant one measures in cosmology is then actually of different origin, in his model due to a GUT scalar field.\\
	In addition, the introduction of $ \tau $ into the action gives light to the fact that this is indeed only \textit{like} a Yang-Mills theory, if even that. If we interpret $ \tau $ as a completely static quantity, it behaves as a Lagrange multiplier. If, however, we give it a kinetic term, we run into the problem of an interpretation, since we are coupling a scalar field to the kinetic term of a supposed spin-2 field. These scalar-tensor theories\cite{ScalarTensor} have existed for about as long as EG, but usually evaded experimental falsifications. Clearly, a more thorough study of the geometrical properties of the MMSW action, such as its connection to Higgs-Chern-Simons forms, will help solving this particular question.\\
	Also, the additional gauge invariance of the theory restricts observables. For one, the Riemann tensor cannot be identified with the $ \mathfrak{so}(1,3) $-part of the field strength, as the latter is a tensor under $ SO(2,3) $. Instead one needs to pick the right combination of $ D\tau,F,... $ to recover standard observables in some gauge. As will be clear later, this actually renders real scalar fields unobservable.

	\chapter[SO(2,3) as a symmetry group]{SO(2,3) as a symmetry group}
\section[Representations of SO(2,3) ]{Representations of SO(2,3) }
	Here, the relevant unitary as well as the finite dimensional representations of $ SO(2,3) $ and its covers $ SP(4) $ and $ L:= \widetilde{SO(2,3)} $ will be presented in a useful form. The material covered here is essentially a mix of works by Evans\cite{EvansIrreps} and Fronsdal, for which there is a review by Nicolai\cite{NicolaiIrreps}, whose conventions will be adopted here. For the finite dimensional reps of the symplectic and orthogonal groups, one may consult Hamermesh\cite{Hamermesh2}, Schwartz\cite{Schwartz} or Weyl\cite{Weyl1997}. 
	
\subsection{The Lie algebra, labels, ladder operators}\label{Lie alg}
	First, a short overview of the Lie algebra of the groups in question is necessary. There will be no distinction between the real and complexified algebra here, since it makes little difference in this thesis.
	The Lie algebra $ \mathfrak{g} = \mathfrak{so}(2,3) $ is given by the matrices \newline $ \{X \in \text{Mat}(5,5) | X^{AB} +X^{BA} = 0 \; X^{AB}= m^A_C \eta^{CB}\} $ skew-symmetric with respect to the standard metric of $ \mathbb{R}^{2,3} $. Note that plain matrices will have mixed component indices here, but apart from this, indices w.r.t the Lie algebra will be all-up for most purposes. 
	The algebra splits into two parts, the Lorentz algebra $ \mathfrak{h} = \mathfrak{so}(1,3) $ and the noncommutative translation subalgebra $ \mathfrak{p} = \mathfrak{g/h} $, so that we write for a basis of the algebra $ \{M_{AB} \} = \{M_{ab}, M_{5b} \} = \{M_{ab}, \Pi_{b} \} $.
	From this split, one can recover the Poincaré algebra via a limiting process called Inonu-Wigner contraction:
	\begin{align}\label{3-1}
		P_a &:= \lim\limits_{\rho \to \infty} \frac{\Pi_{a}}{\rho}\;\\ 
		\; [P_a,P_b] := \lim\limits_{\rho \to \infty} &\frac{1}{\rho^2}{[\Pi_{a},\Pi_{b}]} = \lim\limits_{\rho \to \infty} \frac{1}{\rho^2}{M_{ab}} = 0
	\end{align}
	where one has to think of the $ M_{AB} $ as having the units of $ \hbar $, so that $ P_a $ gains the correct units as well.\\
	The Cartan subalgebra of this is generated by $ M_{12} \text{ and } \Pi_{0} $, which are the generators of the vertical rotation group and the AdS time translations. As such, these can be chosen Hermitean in any representation of the algebra and will be the primary observables of angular momentum and energy. These also generate the $ SO(2)\times SO(3) $ subgroup of $ SO(2,3) $, which is its maximal compact subgroup.
	It also admits subalgebras isomorphic to $ \mathfrak{so}(1,2) $, which are of relevance if one wants to fully classify the states, see Evans\cite{EvansIrreps} for details.
	The algebra representations are labeled by pairs $ (\epsilon,s) $ with $ \epsilon = \epsilon_0 + k $, $ \epsilon_0 \in {[0,1)}\; , \; k \in \mathbb{N}_0 $ and $ 2s \in \mathbb{N}_0 $. $ \epsilon $ labels the lowest energy, while $ s $ the spin, given by the eigenvalue of $ \vec{J}^2 $ as usual. One can thus realise the representations as a sum $V_{\epsilon,s} = \bigoplus_k V_{k,s}$ of representations of $ \mathfrak{su}(2) $ such that the spaces have different, discrete values of energy. These will be glued together by the raising and lowering operators\footnote{In the physics convention with $ M_{AB} = i\, m_{AB} $, this will give the usual commutators again.} \begin{align}\label{3-2}
		\Pi^{\pm}_i = M_{5i}\pm \frac{M_{0i}}{i} \; \text{with}\; 
		{[\Pi_0,\Pi^{\pm}_i]} = \pm iM^{\pm}_i \; \text{and} \; \\
		J^{\pm} = M_{23}\pm \frac{M_{31}}{i} \; \text{with}\; 
		{[M_{12},J^{\pm}]} = \pm iJ^{\pm}
	\end{align}
	which raise and lower $ k $ between the s-spaces and $ m_s $ inside the s-spaces, respectively. As usual, the spin projection values $ m_s $ will run from $ -s \text{ to } s $ in steps of $ 1 $, just like $ k $. Naturally, all these labels are in units where $ \rho = 1$.
	
	One has, as always, a set of Casimir elements which are proportional to the identity on the representation. These are here the quadratic 
	\begin{equation}\label{3-3}
	C_2 = \frac{1}{2}M^{AB}M_{AB} = \Pi_0^2 + \vec{J}^2 - \{ \Pi^{+}_i , \Pi^{-}_i \} \; \text{for any i} \in \{1,2,3\}
	\end{equation} 
	and quartic 
	\begin{equation}\label{3-4}
	C_4 = W_AW^A \, , \; \text{with} \; W_A = \epsilon_{ABCDE} M^{BC} M^{DE}
	\end{equation}
	with $ W $ akin to the Pauli-Lubanski pseudovector, and one can think of it as representing $ W_A = (\vec{J}\cdot\vec{\Pi}, \, \Pi_0 \, \vec{J} - \vec{P}\times\vec{K} ,\, \vec{J}\cdot\vec{K}) $. As such, taking the limit $ \lim\limits_{\rho \to \infty} \frac{1}{\rho^2} C_{2|4}$ will give the Poincaré Casimirs of $ m^2 $ and $ -m^2s(s+1) $. In general, on an $ (q, s) $-representation of the AdS algebra or the groups associated to it, one will have
	\begin{align}\label{3-5}
		C_2 &= q(q-3) + s(s+1)\\  C_4 &= -(q-1)(q-2)s(s+1)
	\end{align} in the sense of the eigenvalue of the operator on the representation. Setting $ q = m\rho $, one recovers the Poincaré expressions from these. However, this does of course not give a one-to-one correspondence as many values of $ q $ will give a similar limit; Take two small values of $ q $ close to each other. If one will result, in the limit, to a mass squared of 0, then both will. This all depends, of course, on the value of $ \rho $, which is hitherto unknown.
\subsection{Unitary representations of the universal cover}\label{Unirreps}	
	Here, I will study projective representations of the AdS isometry algebra. In general, particle states will not be classified according to it, since the base spacetime will not be AdS in the presence of nonzero curvature. In fact, the interpretation of particles becomes questionable in a general curved spacetime. However, as long as one considers perturbations around a background, this is still possible just as over Minkowski space. This way, one can study the particle types involved with MMSW and find representation theoretic justifications to the actions used for them. Of course, as it exists as an internal invariance here, one still needs the projective reps anyway. 
	The possible state spaces of a quantum field theory are classified by representations of its symmetry groups. In particular, the one-particle states of the theory will generate projective unitary irreducible representations (unirreps) of the group, which is at least the spacetime isometry group. So for $ SO(2,3) $ one needs a classification of the unirreps of its covering group $ G $, which descend to $ SO(2,3) $'s projective representations, through Bargmann's theorem.\cite{Bargmann1,Bargmann2} Using those representations, one can construct field operators which correspond to elementary particles. This classification has been done by Evans\cite{EvansIrreps}, who found that the discrete series of $ L $ contracts to the physical representations of the Poincaré group. One finds in this discrete series massive particles, ones with gauge invariance as well as two special representations which deserve a seperate discussion.
	$\mathfrak{so}(2,3), $ in contrast to its sister algebra $  \mathfrak{so}(1,4) $, admits representations such that the minimal energy $ E_0 $ is bounded. These reps lift, uniquely, to the universal cover $ L $, where they become unirreps. The unirreps are labeled by their minimal energy $ E_0 $ and half-integral spin $ s $. As it turns out, there are two inequivalent representations with equal $ |E_0| \text{ and } s $, but with the sign of $ E_0 $ switched, so one with positive and one with negative energies. Here, we only consider the positive energy ones. In these, there are certain restrictions on $ E_0 $ with respect to $ s $ in order to maintain positivity of the inner product. For $ s>=1 $, this is $ E_0 >= s +1 $, while for $ s=0,\frac{1}{2} $, it is $ E_0 >= s + \frac{1}{2} $\cite{EvansIrreps,NicolaiIrreps}. As the external isometry group of AdS, these conditions ensure that the generic loss of unitarity in curved spacetimes is avoided. It seems that a lack of an infrared cutoff would make AdS unsuitable for quantal systems. One can classify the unirreps \textit{D} by values of $ (E_0=q,s) $\footnote{There are many representations here that will have the same Casimir values. They are not unitarily equivalent.}:
	\begin{enumerate}
		\item $ q>s+1 $ for $ s>=1 $, $ q>s+\frac{1}{2} $ for $ s =0,\frac{1}{2} $. These are the generic, massive representations for any spin. Note that low values of $ q $ will seem indistinguishable from the critical value for $ q $ in the $ \rho \to \infty $ limit, but the representations will not admit gauge invariance.  
		
		\item $ q=s+1 $ for $ s>=\frac{1}{2} $, $ q=1,2 $ for $ s = 0 $. These representations can be realised as ones of the \textit{conformal} group $ SO(2,4) $, so that they can be interpreted as properly massless ones. Interestingly, the scalar field is different in this regard as a massless (conformal) scalar field transforms under $ D(1,0)\oplus D(2,0) $. It is therefore in principle a composite particle in AdS. The unirreps for $ s>=1 $ also admit gauge invariance in this case, while the scalar and spinor do not - a result familiar from flat space field theory. Instead, the singletons admit gauge invariance. For this type, the energy spectrum is given by $ E = 1+k+l+s  $ for $ s>0 $, $ E = l + 2k + 1 $ for $ D(1,0) $, $ E = l + 2k + 2 $ for $ D(2,0)$, all for $ k\in \mathbb{R}^+_0, l \in \mathbb{N}_0 $, where $ l $ gives the orbital angular momentum and $ k \in \mathbb{N}_0 $ represents an energy level. Here, one has $ C_2 = 2(s+1)(s-1) = 2(s^2-1)  $, $ C_4 = -s^2(s^2-1) $, so the Casimirs both vanish for $ s = 1 $, while the Casimirs take the same value for both conformal scalar parts.
		
		\item Di $ := D(1,\frac{1}{2}) $ and Rac $:= D(1/2,0) $, the singleton representations\cite{FlatoFronsdalSingleton}, hilariously named after P.A.M. Dirac. When viewed as the conformal group of a 2+1D Minkowski space, these $ SO(2,3) $-unirreps are also a unirrep of $ ISO(1,2) $, namely the only discrete helicity representations. These unirreps are remarkable in that they are basically impossible to detect in one-particle states, as well as the fact that a two-particle state (read: Tensor product) of any combination of Di and Rac will be equivalent to an infinite collection (read: direct sum) of massless particles \textit{of all spins}. In addition, the singletons admit only one free parameter in their energy spectrum, their total angular momentum. Namely, $ E = \frac{1}{2} + l + s $ for Di and Rac, and $ C_2 = -\frac{5}{4} $ as well as $ C_4 = 0 $ for both. As such, as has been pointed out multiple times by Fronsdal, the energy of a singleton can only be measured on cosmological scales. This is supported by the fact that the contraction limit of the singletons is the Poincaré vacuum, so that there is no measurable energy or momentum in them. Singleton physics has been well-studied and is a subtle topic, related to higher-spin theory and conformal field theories. The only known way to couple them is in conformal field theories on the conformal boundary of AdS. This suggests they might be interesting in AdS-CFT correspondence, but pushes them far outside the scope of this thesis. 
	\end{enumerate}
	One may thus impose on a space of functions these representations $ D(q,s) $ by a second order PDE of the form
	\begin{align}\label{3-6} 
		{[C_2 - q(q-3) - s(s+1)]}f_{q,s}(x) = 0
	\end{align} which fixes the first label, as well as subsidiary conditions to make sure it is a spin-$ s $ representation. Here, $ C_2 $ is an appropriate representation of the quadratic Casimir. I can define a representation of the generators on some chart with coordinates $ x^a $,
	\begin{align}\label{3-7}
		M_{ab} = x_ai\partial_b - x_bi\partial_a\\
		M_{5a} = \rho\sqrt{1+\frac{x^ax_a}{\rho^2}} i\partial_a
	\end{align}
	which fulfils the commutation relations of the Lie algebra. This gives 
	\begin{equation}\label{3-8}
		C_2 = -(\rho^2+2x^2)\Box + x^ax^b\partial_a\partial_b + 2x^a\partial_a
	\end{equation}
	so that
	\begin{equation}\label{3-9}
	{[-(1+2\frac{x^2}{\rho^2})\Box + \frac{x^a}{\rho}\frac{x^b}{\rho}\partial_a\partial_b + 2\frac{x^a}{\rho^2}\partial_a - \frac{q}{\rho}\frac{(q-3)}{\rho} - \frac{s(s+1)}{\rho^2}]}f_{q,s}(x) = 0
	\end{equation}
	If one sets $ q = \rho \mu =  \rho\frac{ m c}{\hbar}$ and neglects terms of order $ \frac{1}{\rho} $ and higher, this reduces to
	\begin{equation}\label{3-10}
		{[\Box +\mu^2]}f_{q,s}(x) = 0
	\end{equation}
	so that in the short-scale approximation, AdS-fields correspond to fields of mass $ \mu = \frac{q}{\rho} $. It is noteworthy that even if $ q=3 $, the spin term can give a nonzero effective mass if one goes beyond the approximation. Interestingly, the $ D(3,s) $-unirreps do not exist for $ s > 2 $, so that in a sense, in the short scale approximation, no massless $ s>2 $ fields exist. The corresponding representations would violate the unitarity bound. This is reminiscent of the well-known fact from higher spin theory that one either limits the theory to spins up to two, or has to include all spins.\cite{VasilievTheory} In a sense, there are strict limits to the existence of higher spins and in this case, it means that \textit{effectively massless} representations of spin greater than two of $ SO(2,3) $ do not exist. This does not exclude, however, the existence of conformal gauge fields with $ s>2 $.\\
	In addition, the $ D(2,1) $-irrep for fields is the only conformal one for which the Casimirs vanishes, making it somehow "more massless" than other fields.
	One can thus guess that the well-known fields of the standard model fall into the first class (for matter fields) and the second (for the gauge bosons). So, all force carrier fields should transform under $ D(2,1) $ universally, while the fermions live in $ D(q,\frac{1}{2}) , q>=2$. The case for the Higgs field is not as clear. As was seen in Nicolai\cite{NicolaiIrreps}, one can associate a mass to all of these unirreps. The mass for scalar fields turns out to be able to be negative within a certain range. This suggests a natural range of Higgs masses with which one can perhaps constrain $ \rho $.
\subsection{Finite dimensional reps of $ SO(2,3) $ and $ SP(4) $}\label{finDimirreps}
	Of course, one has to consider how to embed the fields on AdS into more manageable objects. Namely, one wants to use space dependent objects transforming under a non-unitary, finite dimensional representation of $ L $. However, these do not exist as $ L $ is infinite cyclic.\cite{UniCover} One therefore can, at most, consider finite dimensional faithful representations of the double cover of $ SO(2,3) $, which miraculously is the symplectic group $ SP(4; \mathbb{R}) $\cite{VasilievTheory}. The situation is quite similar to the one with $ SO(1,3) $ and $ SL(2;\mathbb{C}) $, where most of the representations are already given by ones of the regular group, while there are some projective finite dimensional ones which exist as regular reps of the double covering. In particular, the Dirac spinor will be realised as the fundamental representation of $ SP(4) $, where it is irreducible. \\
	For low dimensionality, one can generate all but two irreps and the scalar from $ SO(2,3) $'s fundamental irrep \textbf{5}. One proceeds in the usual manner as for the orthogonal group\cite{Weyl1997,Hamermesh2}, by taking tensor products, taking traces with respect to the preserved metric, and thus decomposing each tensor product into traceless parts, which are irreducible under $ SO(2,3) $. So, I get the lowest dimensional irreps with corresponding Young diagrams
	\begin{description}
		\item[\textbf{0}] $ \bullet $ The scalar or trivial irrep, implemented as a function $ \phi $.
		\item[\textbf{5}] \begin{ytableau}
			~      
		\end{ytableau} The fundmental irrep, implemented via $ V^A $.
		\item[\textbf{10}] \begin{ytableau}
			~ & \none   \\
			~ & \none   \\	       
		\end{ytableau}The adjoint representation, used by the connection $ \mathcal{A}^{AB}$.
	
		\item[\textbf{14}] \begin{ytableau}
			~ & ~   \\
			\none & \none   \\	       
		\end{ytableau} Traceless symmetric, for example the 5D metric perturbation $ h_{AB} $.
	\end{description}
	The latter two fulfil $ T^{AB} \eta_{AB} = 0 $ as the condition that projects the tensor square into irreducibles.\\
	This is not possible for the \textbf{4} irrep of $ SP(4;\mathbb{R}) $, which is its fundamental and a projective (or simply two-valued) irrep of $ SO(2,3) $. We give the irreps of it as well, for the lowest dimensions. Lower Latin indices will denote spinor indices here.
	\begin{description}
		
		\item[\textbf{0}] $ \bullet $ The scalar or trivial irrep, implemented as a function $ \phi $. This is the same for both groups.
		\item[\textbf{4}] \begin{ytableau}
			~      
		\end{ytableau} The Dirac spinor $ \Psi^a $, which is the fundamental.
		\item[\textbf{5}] \begin{ytableau}
			~ & \none   \\
			~ & \none   \\	       
		\end{ytableau} The fundamental of $ SO(2,3) $, here expressed as an antisymmetric   tensor $ V^{ab} $
		\item[\textbf{10}] \begin{ytableau}
			~ & ~   \\
			\none & \none   \\	       
		\end{ytableau} The adjoint rep, here as symmetric tensors $ Q_{ab} $.
		\item[\textbf{14}]\begin{ytableau}
			~ & ~   \\
			~ & ~   \\	       
		\end{ytableau} The "Riemann tensor" irrep $ W^{ab}_{cd} $.
		\item[\textbf{16}] \begin{ytableau}
			~ & ~   \\
			~ & \none   \\	       
		\end{ytableau} The second double-valued irrep, realised in $ Z^{ab}_c $. This likely corresponds to the Rarita-Schwinger spinor, which can be realised by taking the latter as $ Z^a_\mu $, with $ 16 $ components as it transforms under $ \left(1,\frac{1}{2}\right) \oplus \left(\frac{1}{2},1\right) $.
	\end{description}
	These, unlike their $ SO(2,3) $ counterparts, are not traceless with respect to the 5D metric, but an antisymmetric inner product defined in the fundamental as \begin{align}\label{3-11}
	C_{ab} = -\varepsilon_{ba} = \begin{tabular}{|c|c|c|c|}
	\hline 
	0 & -1 & 0 & 0 \\ 
	\hline 
	1 & 0 & 0 & 0 \\ 
	\hline 
	0 & 0 & 0 & 1 \\ 
	\hline 
	0 & 0 & -1 & 0 \\ 
	\hline 
	\end{tabular} = \begin{tabular}{|c|c|}
	\hline 
	$ \epsilon $ & 0 \\ 
	\hline 
	0 & $ -\epsilon $ \\ 
	\hline 
	\end{tabular}
	\end{align}
	with $ \epsilon $ the usual antisymmetric symbol in 2D. This inner product matrix can also be written using the Dirac matrices as $ C = -i\gamma^0\gamma^2 $. So, all but the \textbf{4} and \textbf{16} irreps can be related to each other. For example, \begin{equation}\label{3-12}
		V^A = \phi^T C \gamma^A \chi \;\text{, with } \phi,\chi \in \textbf{4} \text{ and } \phi^T C \chi = 0
	\end{equation}Since $ C $ is antisymmetric, the $ 16 $ components of $\phi,\chi $ become $ 6 $ free ones, of which another one is projected out so that this expression has $ 5 $ degrees of freedom. One can check by using standard identities that this indeed transforms as a \textbf{5}-vector.\\
	One can reduce these representations to ones of $ SL(2,\mathbb{C}) $ and $ SU(2) $. Since this is significant for this work, some of the decompositions are given here. The notation is as follows: \\
	$ SO(2,3)/SP(4)$-irrep $ \simeq SL(2,\mathbb{C}) $-irreps denoted by spin pairs  $ \simeq SU(2) $-irreps denoted by \textbf{spin}.
	\begin{description}
		\item[\textbf{0}] $ \simeq  \left(0,0\right) \simeq \textbf{0}$ 
		\item[\textbf{4}] $ \simeq  \left(\frac{1}{2},0\right)\oplus\left(0,\frac{1}{2}\right) \simeq \mathbf{\frac{1}{2}} \oplus \bar{\mathbf{\frac{1}{2}}}$
		\item[\textbf{5}] $ \simeq  \left(\frac{1}{2},\frac{1}{2}\right)\oplus\left(0,0\right) \simeq \textbf{1}\oplus\textbf{0}\oplus\textbf{0}$
		\item[\textbf{10}] $ \simeq  \left( \left(1,0\right)\oplus\left(0,1\right) \right) \oplus \left(\frac{1}{2},\frac{1}{2}\right) \simeq \textbf{1}\oplus\textbf{1}\oplus\textbf{1}\oplus \textbf{0} $
		\item[\textbf{14}] $ \simeq  \left(1,1\right)\oplus\left(\frac{1}{2},\frac{1}{2}\right) \oplus\left(0,0\right) \simeq \textbf{2} \oplus \textbf{1} \oplus \textbf{0} \oplus \textbf{1} \oplus \textbf{0} \oplus \textbf{0}$
	\end{description}

	A few remarks about the reps of low dimension are in order. While they do depend on $ q$, they do so in the most trivial manner: for given spin $s \text{ in } (q,s) $, the representations will have the same dimension, but transform differently under time translations. Namely, just like the irreps of $ SO(2) $ are all 2-dimensional, but differ in rotation speed, which reduces the group composition to addition of real numbers, one gets in a generic exponential element
	\begin{equation}\label{3-13}
		S = \text{exp}(-\frac{i}{2}\epsilon^{AB}\pi(M_{AB})) = \text{exp}(-f(q)\frac{i}{2}\epsilon^{5a}\pi(M_{5a})  -\frac{i}{2}\epsilon^{ab}\pi(M_{ab}))
	\end{equation}
	If one restricts to the $ SO(2) $-subgroup, the extra factor in the translation part will ensure that one restricts to a $ q $-representation of it. In fact, without introducing that factor, one will always seem to find $ q=s $ for the finite dimensional representations, though there is, to my knowledge, no proof of such a statement.
	One may restrict a representation of $ SP(4) $ to its maximal compact subgroup $ SO(2)\times SU(2) $, which has finite dimensional representations labeled by energy and spin. This makes it possible to identify the $ (q,s) $-values of these finite dimensional reps. Take, for example, the fundamental of $ SP(4) $, the Dirac spinor rep. A general element near the identity of the group is represented by 
	\begin{equation}\label{3-14}
		S = \text{exp}(-\frac{i}{2}\epsilon^{AB}\Sigma_{AB}) = \text{exp}(-\frac{i}{2}x^{a}\gamma_{a}  -\frac{i}{2}\epsilon^{ab}\Sigma_{ab})
	\end{equation}
	setting $ x^i, \epsilon^{0i} = 0 $ gives the restriction
	\begin{equation}\label{3-15}
	S = \text{exp}(-\frac{i}{2}t\gamma_{0}  -\frac{i}{2}\epsilon^{ij}\Sigma_{ij})\\
	= \text{exp}(-\frac{i}{2}t\gamma_{0})  \text{exp}(-\frac{i}{2}\epsilon^{ij}\Sigma_{ij})
	\end{equation}because the rotation and time generators commute.
	Of course, the right part is just the usual Dirac bispinor representation of the rotation group. Disregarding it, one finds for time translations
	\begin{equation}\label{3-16}
	S = \text{exp}(-\frac{i}{2}t\gamma_{0}) = \text{cos}(\frac{t}{2}) \textbf{1} - i\text{sin}(\frac{t}{2})\gamma_{0}
	\end{equation}so that one recovers a representation of $ SO(2) $ corresponding to $ q = \frac{1}{2} = s $. To gain a representation with $ q=1 $, one will have to introduce a factor for the $ \gamma_{0} $-part. Similarly, the fundamental of $ SO(2,3) $ gives rise to a $ q=1=s $-rep of $ SO(2) $.
	\newpage
\section[Primary considerations for actions]{Primary considerations for actions}\label{PrimCons}
	Here I give my own thoughts about generally acceptable actions in the context of MMSW.
	First, let me propose a few conditions on actions in order to be compatible with the principles of the MMSW formalism. These are similar to Fukuyama's conditions.\cite{Fukuyama}
	\begin{enumerate}
		\item Diffeomorphism invariance.
		\item Metric independence.
		\item Local SO(2,3)-invariance.
		\item Reduction to local $ SO(1,3) $-invariance once the standard gauge is assumed for $ \tau $.
		\item Reduction to the standard actions in the standard gauge and under Inönü-Wigner contraction of the representations.
	\end{enumerate} 
	In addition, only renormalisable couplings will be considered here for simplicity, and, of course, actions must be real-valued.\\
	1) means that, as is usus in other theories, the spacetime manifold itself is devoid of physical content. Any two manifolds which admit the same smooth structure are indistinguishable and, in particular, the physics is independent of chosen coordinates.\\
	2) is given by the fact that a theory of MMSW type must work purely in terms of the principal connection on $ Q $ before the vierbein acquires its standard role. Otherwise, inconsistencies occur. This condition is equivalent to the actions being those of a coupled topological theory. This is a very strong requirement. However, this does not mean that the EM tensor of the theory vanishes as it is now formulated through variation with respect to the MMSW connection.\\
	4) This is self-explanatory. \\
	5) This means that up to gauge fixing removing nonstandard terms in the action, the coupling to other fields must generate the canonical volume element, if needed, and at most generate terms which scale as $ \mathcal{O}(\frac{1}{\rho^2}) $ or similar.
	
	Generally, one will have to use the covariant derivative of the tangency map to couple properly. This is due to its property that, in the standard gauge\cite{Fukuyama},
	\begin{align}\label{3-17}
		D\tau^A &= d\tau^A + \mathcal{A}^{AB}\tau_B\\
		&\to  \mathcal{A}^{A5} = -\frac{\rho}{\rho} \theta^a \delta^A_a
	\end{align}
	So, I can incorporate this in order to generate a volume form. Generically, for $ {[f]} = 4 $, \begin{align}\label{3-18}
		\begin{split}
		S_f &= \frac{1}{4!\rho}\int_{M}  \; f \; \circledast(D\tau\wedge D\tau\wedge D\tau\wedge D\tau) \\
		&= \frac{1}{4!\rho}\int_{M} \epsilon_{ABCDE} \; f \; D\tau^A\wedge D\tau^B\wedge D\tau^C\wedge D\tau^D \tau^E \\
		&\to \frac{1}{4!}\int_{M} \epsilon_{abcd} \; f \; \theta^a\wedge \theta^b\wedge \theta^c\wedge \theta^d\\
		&= \int_{M} vol \; f
		\end{split}
	\end{align} in the standard gauge, so one can create any scalar (non-derivative) term with this action that will reduce to the standard integral for functions. This will not work for derivatives, though. \\
	A second way in which one can introduce couplings that do not contain derivatives in the unbroken phase is to use an action of the type
	\begin{align}\label{3-19}
		S_5 &= \frac{1}{4!} \int_M Q^A D\tau^B\wedge D\tau^C\wedge D\tau^D\wedge D\tau^E \epsilon_{ABCDE} \\
		&\to \int_M vol \, Q^5
	\end{align} which, in the standard gauge, shows that only a Lorentz scalar part of the \textbf{5}-vector $ Q^A $ contributes to the action. I will later use this to create Yukawa couplings.
	Overall, one will need to create a metric-independent, SO(2,3)-invariant four-form to integrate over from just the derivatives of the fields, $ \tau $ and $ D\tau $, which guarantees the conditions above. Note that $ {[\tau^A]} = -1 = {[D\tau^A]} $, so that $ {[Q^A]} = 4 $.\\
	One will, for representation theoretic purposes, have to use the transformation properties under diffeomorphisms in addition to the internal representation to show the particle content of an action.
	For example, taking the regular Palatini formalism, with Minkowski space as a background, one can take the four-vector-valued one-form $ V^a_\nu $. It will transform, under combined internal Lorentz and external Poincaré transformations, as 
	\begin{align}\label{3-20}
	\stackrel{\stackrel{\begin{ytableau}~\end{ytableau}}{\otimes}}{\begin{ytableau}~\end{ytableau}}
	\, \simeq \begin{ytableau}~\end{ytableau}\otimes\begin{ytableau}~\end{ytableau}\simeq \begin{ytableau}
	~ & ~\\
	\none & \none\\
	\end{ytableau}\oplus\begin{ytableau}
	~ & \none\\
	~ & \none\\
	\end{ytableau}\oplus \bullet
	\end{align} 
	Here, the upper part shows the representation associated to the internal index, while the lower part shows the spacetime representation. The far right hand side shows the reduced Lorentz representation the object transforms in. As such, this one-form can actually describe $ s=2,1,0 $-fields, even though a first look suggests otherwise. One can turn the object into a totally internal two-tensor by use of $ \eta,g,\theta $ and $ e $ when available. Then, the way it transforms becomes more transparent, but this is not possible for internal $ SO(2,3) $-reps.\\
	The equivalent procedure for these representations is to first decompose them into $ SO(1,3) $-reps as given in section \ref{finDimirreps}. Then, for example for an object $ V^A_\mu = (V^a_\mu, V^5_\mu) $, the full representation under an external and internal Lorentz group is given as 
	\begin{align}\label{3-21}
		\stackrel{\stackrel{\begin{ytableau}
			~ 	
		\end{ytableau}}{\otimes}}{\left(\frac{1}{2},\frac{1}{2}\right)}
	\simeq 
		\stackrel{\stackrel{\left(\frac{1}{2},\frac{1}{2}\right)\oplus \left(0,0\right)}{\otimes}}{\left(\frac{1}{2},\frac{1}{2}\right)}
	\simeq
		\left(\frac{1}{2},\frac{1}{2}\right)\otimes \left(\frac{1}{2},\frac{1}{2}\right) \oplus \left(\frac{1}{2},\frac{1}{2}\right)\\
	\simeq 
		\left(1,1\right) \oplus \left(1,0\right) \oplus \left(0,1\right) \oplus \left(\frac{1}{2},\frac{1}{2}\right) \oplus \left(0,0\right)
	\end{align} with Young tableaux now given for $ SO(2,3) $-irreps. So one can see that one can embed $ s=2,1,0 $ fields in this type of object, yet again.
	 \\
	It is instructive to consider the MMSW field for a moment. It is described by an object $ A^{AB}_\mu   = (A^{ab}_{\mu}, A^{5b}_{\mu})$ which thus transforms under the Lorentz group irrep 
	\begin{align}
		(\left(1,0\right) \oplus \left(0,1\right) \oplus \left(0,0\right)) \otimes\left(\frac{1}{2},\frac{1}{2}\right)
	\end{align}
	\begin{align}\label{3-22}
		\simeq \left[ \left(\frac{1}{2},\frac{1}{2}\right) \oplus \left(\frac{3}{2},\frac{3}{2}\right) \oplus \left(\frac{1}{2},\frac{1}{2}\right)  \right]\oplus \left[ \left(1,1\right) \oplus \left(1,0\right) \oplus \left(0,1\right) \oplus \left(0,0\right) \right] 
	\end{align} where the first $ \left(\frac{1}{2},\frac{1}{2}\right) $-irrep is a pseudovector under reflections, which are not considered here. The first square brackets contain the irreps corresponding to the spin connection part $ A^{ab}_{\mu} $, the second ones contain the vierbein $ A^{5b}_{\mu} $. On-shell, the vierbein determines the spin connection through the torsion freeness condition, so that in this reduced action, one has only the second part. Of course, this analysis only makes sense in the broken phase, when the vierbein exists. There, one can switch between internal Lorentz and spacetime indices freely in order to give the field content an interpretation in terms of irreps. Say one wants to remove the spin-1 and spin-0 parts of $ \theta $, which corresponds to
	\begin{align}\label{3-23}
		&\theta^{ab} = \theta^{ba} & \theta^{ab} \eta_{ab} = 0 & 
	\end{align}where $ \theta^{ab} : = \theta^{a}_\mu e^\mu_c \eta^{bc} $. Then, there are 9 degrees of freedom left, which give the metric constructed from $ \theta $ 9 DoF. One can thus see that this gives all freedom of a traceless perturbation of the background metric, which is the potential used for linearised gravity. In the original form though, one has in principle $ 16 $ degrees of freedom for the vierbein. $ 6 $ of these can be taken as gauge due to internal Lorentz invariance. Similarly, diffeomorphisms are able to remove four more degrees of freedom, with constraints taking away four more, leaving the two of the propagating graviton.

	\chapter[Actions and couplings in MMSW]{Actions and couplings with gauged SO(2,3)-invariance}

	Here, I will propose and discuss kinetic terms for some standard fields and couplings between them. Many results here relate to work by Fukuyama\cite{Fukuyama}.

\section[Dirac spinors]{Dirac spinors}\label{Dirac}
	The Dirac spinor is possibly the most simple case of a coupling and has been discussed by Fukuyama 35 years prior to this thesis. In my convention, his action reads
	\begin{align}\label{4-1}
		S_{Fukuyama} &= \frac{-1}{3!}\int_{M} \epsilon_{ABCDE} \; \bar{\psi}\left(  i\Sigma^{AB} D + \frac{\lambda\rho}{4} \tau^A D\tau^B \right) \psi \, \wedge D\tau^C\wedge D\tau^D \wedge D\tau^E  \\
		&= \frac{1}{3!}\int_{M} \; \bar{\psi}(D\tau^3)_{AB}\Sigma^{AB} \wedge iD\psi \, + \frac{-\lambda}{4!\rho} \int_{M} \bar{\psi}\psi \circledast(D\tau^4)
	\end{align}with the covariant derivative 
	\begin{align}\label{4-2}
		D\psi &= d\psi + i\frac{1}{2}\mathcal{A}^{AB} \Sigma_{AB} \psi\\
		 &= d\psi + is\frac{1}{2}\omega^{ab} \Sigma_{ab} \psi + i\frac{q}{2\rho} \theta^a \gamma_{a} \psi
	\end{align}in the \textbf{4}-irrep of $ SP(4) $, and $ {[\lambda]} = +1 $. This action is, by transformation properties of $ \psi $ and $ \Sigma $, manifestly diffeomorphism and $ SO(2,3) $ invariant. In the standard gauge, 
	\begin{align}\label{4-3}
		\begin{split}
		S_{StG} &= \frac{-2}{3!}\int_{M} \epsilon_{fbcd} \; \bar{\psi}  \Sigma^{5f} iD \psi \, \wedge D\tau^b\wedge D\tau^c \wedge D\tau^d - \lambda \int_{M} vol  \bar{\psi}\psi \\
		&= \frac{2}{3!}\int_{M} \epsilon_{fbcd} \; \bar{\psi} \Sigma^{5f} iD_\alpha  \psi \,e^\alpha_a \theta^a\wedge \theta^b\wedge \theta^c \wedge \theta^d - \lambda \int_{M} vol  \bar{\psi}\psi\\
		&= 2\int_{M} vol \; \delta^a_f \; \bar{\psi}  \Sigma^{5f} iD_\alpha \psi \,e^\alpha_a  +- \lambda \int_{M} vol  \bar{\psi}\psi \\
		&= \int_{M} vol \; \left( \bar{\psi} \gamma^{b} \,e^\alpha_b iD_\alpha \psi   - \lambda \bar{\psi}\psi \right) = \int_{M} vol \; \bar{\psi}\left( i\slashed{D}  - \lambda \textbf{1}\right)\psi
		\end{split}
	\end{align}While this superficially looks like the Dirac action for curved spacetime, it still contains the $ \theta $-part of the $ SO(2,3) $-connection. One can decompose it thus, using $ D\psi = d\psi +i\frac{1}{2}\mathcal{A}^{AB}\Sigma_{AB}\psi = d\psi +i\frac{s}{2}\mathcal{A}^{ab}\Sigma_{ab}\psi + iq\mathcal{A}^{5b}\Sigma_{5b}\psi$:
	\begin{align}\label{4-4}
		i\gamma^ae^\mu_aD_\mu\psi &=  i\gamma^ae^\mu_a\dot{D}_\mu\psi - \frac{q}{2}\gamma^ae^\mu_a\mathcal{A}_\mu^{5b} \gamma_{b}\psi\\
		&= i\gamma^ae^\mu_a\dot{D}_\mu\psi - \frac{q}{2\rho}\gamma^ae^\mu_a\theta^{b}_\mu \gamma_{b}\psi\\
		&=  i\gamma^ae^\mu_a\dot{D}_\mu\psi - \frac{2q}{\rho}\psi
	\end{align} with $ \dot{D} $ the covariant derivative w.r.t. the $ SO(1,3) $-connection. This gives the action
	\begin{align}\label{4-5}
		S = \int_{M} vol \; \bar{\psi}\left( i\slashed{\dot{D}}  - (\lambda + \frac{2q}{\rho}) \textbf{1}\right)\psi 
	\end{align}This shows that even in the absence of a quadratic coupling, the action still displays a positive mass term. This works as a natural infrared regulator for all fermions. In fact, if one removes the quadratic coupling and makes the coupling to the vierbein variable take the form $ D\psi = \dot{D}\psi + q\mathcal{A}^{5b}\Sigma_{5b}\psi $, thus making it transform in the $ (q,\frac{1}{2}) $-irrep of $ SP(4) $, one has an effective mass term of value $ m_{AdS} = \frac{2q}{\rho} $. Thus, for the ($ q=1 $), one would have the minimal value for the mass, while $ q>1 $ it is unbounded above. The interpretation is that there is a maximal correlation length $ \xi = \frac{1}{m_{AdS}} = \frac{\rho}{2} $ for spin-$ \frac{1}{2} $ fields with internal $ SO(2,3) $-invariance. The natural mass term is thus completely encoded in the internal transformation properties of the fields. Further quadratic couplings will, of course, introduce changes to the mass.\\
	 \\
	Before considering Majorana spinors, I note the two bilinear products on \textbf{4} which are invariant.
	\begin{align}\label{4-6}
		B(\phi,\chi) &= \phi^\dagger \gamma^{0} \chi = \bar{\phi}\chi\\
		M(\phi,\chi) &= -i\phi^T\gamma^{0}\gamma^{2} \chi = \phi^T C \chi
	\end{align}These are the only real, invariant bilinear forms in this representation.
	Thus the unique scalar one-form terms are
	\begin{align}\label{4-7}
		B(\psi,D\psi) &= \bar{\psi} D\psi & M(\psi, D\psi) &= \psi^T C\, D\psi
	\end{align}One can get more interesting products by introducing Clifford algebra elements, which all have definite transformation properties:
	\begin{align}\label{4-8}
		&B(\psi,\gamma^AD\psi) = \bar{\psi}\gamma^AD\psi &  B(\psi,\Sigma^{AB}D\psi)= \bar{\psi}\Sigma^{AB}D\psi \\
		&M(\psi,\gamma^AD\psi)= \psi^T C\gamma^AD\psi & M(\psi,\Sigma^{AB}D\psi) = \psi^T C\Sigma^{AB}D\psi
	\end{align} which transform according to their indices as tensors of the \textbf{5} irrep as expected. Of course, higher order Clifford terms are also allowed, but they will not be relevant here. A simple check shows that both $ \bar{\psi}\gamma^AD\psi $ as well as $  \bar{\psi}\Sigma^{AB}D\psi $ give rise to the same action as Fukuyama's in the standard gauge. So both are legitimate actions giving rise to massive spinors.\\
	Here, I will prefer the variant with one gamma matrix due to its simplicity, as well as the change one has to make to the Fukuyama action due to it:
	\begin{align}\label{4-9}
	\begin{split}
	S &= \frac{-1}{3! \rho}\int_{M}  \; \bar{\psi}i \gamma^A D\psi \, \wedge \circledast (D\tau \wedge D\tau \wedge D\tau)_A\\
	&= \frac{-1}{3!\rho}\int_{M} \epsilon_{ABCDE} \; \bar{\psi}i \gamma^A D\psi \, \wedge D\tau^B\wedge D\tau^C \wedge D\tau^D \tau^E\\
	&= \frac{1}{3!\rho}\int_{M}  \; \bar{\psi}i \gamma^A\circledast ((D\tau)^3)_A\wedge D\psi \, 
	\end{split} 
	\end{align}
	If one wants to use chiral spinors, the equivalent of Weyl spinors, or Majorana spinors, one has to impose constraints on $ \psi $:\cite{FukSpinor}
	\begin{description}
		\item[Weyl]    $ \mathcal{P}^{\pm}\psi = \psi $
		\item[Majorana] $ (\bar{\psi}C)^T = i\gamma^{0}\gamma^{2}\gamma^{0}\psi^{\ast} = -i\gamma^{2}\psi^{\ast} = \psi $
	\end{description}
	The second condition implies $ M(\psi^\ast,\psi^\ast) = B(\psi,\psi) $, which one could interpret as the spinor being real.\\
	Each of the actions here gives a spin-$ \frac{1}{2} $ field, as one only has internal transformations to worry about.
	\newpage
\section[Scalar fields]{Scalar fields}\label{Scalar}
	Some of the generic problems of writing metric-independent actions already appear in the case of the scalar particle. For scalars, any scalar product will give rise to an $ SO(2,3) $-invariant bilinear, so that for a single complex scalar, one can only have 
	\begin{align}\label{4-10}
		\phi^\dagger\phi , (\phi^\dagger\phi)^2 \; ,\; \phi^\dagger D\phi + h.c. , (D\phi)^\dagger \wedge D\phi  
	\end{align} as renormalisable terms. While the former two can be subsumed in the standard form (\ref{3-18}), the latter two are candidates for kinetic terms. However, in the scalar representation, there are no objects like the gamma matrices, which allow to couple to $ \tau^A $.\\ 
	Namely, one could use an action of the type
	\begin{align}\label{4-11}
		S &= \frac{1}{\rho}\int_M ((D\phi)^\dagger \wedge D\phi)^{AB}\wedge \circledast(D\tau \wedge D\tau)_{AB}\\
		&= \frac{1}{2!}\int_M vol \, ((D_\mu\phi)^\dagger D_\nu\phi)^{ab}\,  (e^\mu_a e^\nu_b - e^\mu_b e^\nu_a)\\
		&=\int_M vol \, ((D_\mu\phi)^\dagger D_\nu\phi)^{\mu\nu}
	\end{align}which does not produce a scalar field action as $ M $ is antisymmetric. Thus, one needs a more elaborate construction. One, also due to Fukuyama\cite{Fukuyama}, is to use a first-order action for a $ \textbf{5} $-vector $ \phi^A $:
	\begin{align}\label{4-12}
		\begin{split}
		S &= \frac{1}{3!\rho} \int_M \epsilon_{ABCDE} \; \phi^A D\phi^B\wedge D\tau^C\wedge D\tau^D\wedge D\tau^E \\
		&= \frac{-1}{3!\rho}\int_M \epsilon_{abcd} (\phi^5 D\phi^a - \phi^a D\phi^5) \wedge \theta^b \wedge \theta^c \wedge \theta^d\\
		&= \frac{-1}{3!\rho}\int_M  vol \, \epsilon^{rbcd} \epsilon_{abcd}  \left( \phi^5 D_r\phi^a - \phi^a D_r\phi^5 \right) \\
		&= \frac{-1}{\rho}\int_M  vol \, e^\mu_a  \left( \phi^5 D_\mu\phi^a - \phi^a D_\mu\phi^5 \right)
		\end{split}
	\end{align} after assuming the standard gauge again. This action can be written as, using $ M_{AB}^\mu\text{, with }M_{a5}^\mu = e^\mu_a \text{,but otherwise } 0 $ and antisymmetric,
	\begin{align}\label{4-13}
		S &= \frac{1}{\rho}\int_M  vol \, \phi^A M_{AB}^\mu D_\mu \phi^B
	\end{align}which has the following equations of motion:
	\begin{align}\label{4-14}
		D_\mu (\phi^A M_{AB}^\mu ) &= M_{BA}^\mu ( D_\mu\phi)^A  \\
		\;  2 D_\mu(\phi^A M_{AB}^\mu) &= D_\mu(M_{AB}^\mu)\phi^A
	\end{align}where the free index $ B $ on the LHS is understood to be outside the bracket, so that this is not just a partial derivative.
	\begin{align}\label{4-15}
		& 2 D_\mu(\phi^5 e_a^\mu) = D_\mu(e_a^\mu) \phi^5 & 2 D_\mu(\phi^a e_a^\mu) = D_\mu(e_a^\mu) \phi^a &
	\end{align} To see where this is heading, take the case where $ D_\mu(e_a^\mu) = 0 $, which, in the equivalent Riemannian case, corresponds to
	\begin{align}\label{4-15a}
		 \Gamma^\rho_{\rho \mu} = 0 = \partial_\mu(\sqrt{-g})) 
	\end{align}
	In this case, which assumes the volume element to be constant, the equations simplify to
	\begin{align}\label{4-16}
		& (D_\mu\phi)^5 = 0 & e^\mu_a (D_\mu\phi)^a = 0&
	\end{align}
	Using the first relation, one can find $ \phi^a = -\rho\eta^{ab}e^\mu_b\partial_{\mu}\phi^5 $. As such, the $ \phi^a $ are identified as the derivatives of the field $ \phi^5 $ and this yields a second order equation from the second equation of motion:
	\begin{align}\label{4-17}
		\begin{split}
		&e^\mu_a (D_\mu\phi)^a = e^\mu_a (\partial_{\mu}(\phi^a) + \omega^a_{\mu b} \phi^b + \mathcal{A}^a_{\mu5}\phi^5)\\
		&= -\rho^2e^\mu_a (\partial_{\mu}(\eta^{ac} e^\nu_c\partial_{\nu}\phi^5) + \omega^a_{\mu b}\eta^{bc} e^\nu_c\partial_{\nu}\phi^5  +\frac{1}{\rho^2}\theta^a_{\mu}\phi^5)\\
		&= -\rho^2e^\mu_a (\eta^{ac}\partial_\mu(e^\nu_c)\partial_\nu(\phi^5) + \dot{D}(e^\nu)_c \eta^{ac}\partial_\mu(\phi^5)  +\frac{1}{\rho^2}\theta^a_{\mu}\phi^5)\\
		&= -\rho^2 (e^\mu_a\eta^{ac}e^\nu_c\partial_\mu(\partial_\nu(\phi^5)) + e^\mu_a\dot{D}_\mu(e^\nu)_c \eta^{ac}\partial_\mu(\phi^5)  +\frac{1}{\rho^2}e^\mu_a\theta^a_{\mu}\phi^5)\\
		&= -\rho^2(g^{\mu \nu} \partial_\mu \partial_\nu \phi^5 + \frac{4}{\rho^2}\phi^5 ) = 0 \\
		&\Rightarrow (\square + m_{AdS}^2 )\phi^5 = 0 \qquad \qquad m_{AdS} = \frac{2}{\rho}
		\end{split}
	\end{align} where in the last line the condition on the affine connection was used again.
	One can enforce the unirrep of the embedded spin-0 field by setting the spin-connection coupling to zero:
	\begin{align}\label{4-18}
		(D\phi)^a &= d\phi^a + s \, \omega^{ab}\phi_b - \frac{q}{\rho}\theta^b \phi_5\\
		(D\phi)^5 &= d\phi^5 + \frac{q}{\rho}\theta^b \phi_b
	\end{align}
	so the $ 5 $th component does not have a spin-part, as is to be expected. By setting $ s=0 $, one should acquire a description of a spin-$ 0 $ field, though a check of this will be left to future investigations. While one could set $ q=1,2 $ to make a conformal scalar, there is no guarantee for the action to actually be conformally invariant.\\
	One may also use Fukuyama's action\cite{Fukuyama}
	\begin{align}\label{4-19}
		\begin{split}
		S_{conf} &= -\frac{1}{2\rho}\int_{M} (\phi^F\tau_F)^2 \, F^{AB}\wedge \circledast(D\tau^2)_{AB} \\
		&\to -\rho^2\int_{M} F^{ab}\wedge \theta^c\wedge \theta^d \epsilon_{abcd} (\phi^5)^2\\
		&= -\rho^2\int_{M} vol  \, (R +\frac{12}{\rho^2})(\phi^5)^2
		\end{split}
	\end{align}
	which describes a conformal scalar.

	One sees that a working method for embedding scalars was to introduce them as a component of the \textbf{5}-irrep. Since  $\mathbf{5} \simeq  \left(\frac{1}{2},\frac{1}{2}\right)\oplus\left(0,0\right)$, the derivative part, which is the $ 4 $-vector, decouples from the 5th component, which then transforms as a scalar under the residual internal $ SO(1,3) $-invariance. Namely, under an Anti-deSitter transformation, \begin{align}\label{4-20}
		\phi(x)^A &\to \Lambda(x)^A_B \phi(x)^B\\
		(\phi^a,\phi^5) &\to (\Lambda^a_b\phi^b + \Lambda^a_5 \phi^5 , \, \Lambda^5_b \phi^b + \Lambda^5_5 \phi^5)
	\end{align}for example under $ \Lambda^a_b = \delta^a_b, \Lambda^5_5 = \Lambda^0_0 = \cos(\alpha), \Lambda^5_0 = -\sin(\alpha) = -\Lambda^0_5 $, which is time translation with $ \Delta t = \frac{\rho \alpha}{c} $:
	\begin{align}\label{4-21}
	(\phi^0,\phi^i,\phi^5) &\to (\Lambda^a_b\phi^b + \Lambda^a_5 \phi^5 , \, \Lambda^5_b \phi^b + \Lambda^5_5 \phi^5)
	\end{align}so that, seemingly, the scalar and its time derivative mix. However, this is not an issue. Since one can freely choose any gauge, it is clear that the dynamics cannot actually depend on the section chosen. As such, one will always find a way to eliminate four components of the $ 5 $-vector involved in favour of the remaining component, which will become the propagating scalar.\\
	Without the restriction above, one instead finds 
	\begin{equation}\label{4-22}
	 \phi^a = -\eta^{ab} \rho \left( \partial_\mu(\phi^5) e^\mu_b  + \frac{1}{2} \phi^5 \partial_\mu (e^\mu_b) )\right) 
	\end{equation}Using this in the last line of (\ref{4-12}) yields 
	\begin{align}\label{4-23}
		S =\int_M vol \,  \left[ e^\mu_a \phi^a \partial_\mu(\phi^5) + \frac{1}{\rho} \phi^a\phi_a + \phi^5 (\dot{D}_\mu \phi)^a e^\mu_a - \frac{4}{\rho} (\phi^5)^2 \right]
	\end{align}which, in turn, results in the hefty expression
	\begin{align}\label{4-24}
		\int_{M} vol (\, 3 \partial_{\mu}\phi \partial^{\mu}\phi + \phi \square \phi -\frac{1}{2}(\frac{2}{\rho})^2\phi^2 + \frac{1}{4}\phi^2 \partial_{\alpha}e^{\alpha a} \partial_{\sigma}e^{\sigma}_a \\
		+ \frac{1}{2} \phi e^\mu_a \dot{D}_\mu (\partial_{\nu}e^\nu)^a + \frac{5}{2} \phi \partial^{\mu}\phi \theta^b_\mu \partial_{\nu} e^\nu_b + \phi \partial_\nu \phi e^\mu_a (\dot{D}_\mu e^\nu)^a)
	\end{align}upon inserting (\ref{4-22}). Here, the distinction between $ \theta $ and $ e $ has been neglected for simplicity, as one is in the broken phase for this calculation.\\
	This action gives rise to the equation of motion:
	\begin{align}\label{4-25}
		&\left[\square + \left(\frac{2}{\rho}\right)^2\right] \phi - \frac{1}{8}e^\mu_a \dot{D}_\mu(\partial_{\nu}e^\nu)^a \\
		&+\phi \left(\partial_{\alpha}e^{\alpha a} \partial_{\sigma}e^{\sigma}_a + \frac{5}{8}\theta^{b\mu}\partial_{\mu}\partial_{\nu}e^\nu_b  + \frac{1}{4}\partial_{\mu}(e^\nu_a (\dot{D}_\nu e^\mu)^a)  \right) = 0
	\end{align} These seem to not reduce directly to the last line of  (\ref{4-17}) when the constraint (\ref{4-15a}) is imposed. However, the leading term that does not depend on derivatives of $ e $ is the same as for a free scalar field. Since the full equations include the coupling to the MMSW field, this extension is to be expected. However, it is unknown to me how one has to impose the constraint on these full equations of motion to arrive at the expected equations in terms of the affine connection corresponding to $ \mathcal{A} $. As Fukuyama mentioned, but did not explain in detail, the leading order terms of this equation should give rise to the standard coupling of a scalar to gravity. Thus, it seems that in the MMSW formalism, scalar fields are predicted to have additional interaction terms with the gravitational fields, which might be falsifiable. However, lacking solutions to these equations, one can only speculate.\\
	To see why one needs a nontrivial, finite representation to embed the scalars, one only has to look to other gauge theories. If a field transforms under an internal invariance group, the corresponding covariant derivative will need to include generators in a fitting, finite representation of the group. For a nonabelian group, there are no nontrivial one-dimensional, finite dimensional representations, so one cannot use 1D irreps to embed fields charged under the internal invariance. For example, in $ SU(2) $ Yang Mills theory, one cannot describe a weakly charged scalar singlet - one needs a multiplet that transforms nontrivially (in the \textbf{2}-irrep) under the internal group. Analogously, for the $ SO(2,3) $-invariance of MMSW, this implies that fields with spin, \textit{even scalar fields}, transform in a nontrivial finite-dimensional irrep of $ SP(4) $. \\
	Thus, one can see this formulation of the scalar field as the most efficient one, but one might probably be able to use the \textbf{4}-irrep as well. In addition, this argument shows that in general, objects which transform trivially under the internal $ SO(2,3) $ will lack an interpretation in terms of particle carriers. Here I only considered real scalars. If one used more complicated representations, the situation might change enough to grant a description which is less nonlinear.\\
	There is a final and severe problem for this construction: The component identified with the propagating scalar is not gauge invariant - it is part of a multiplet - it cannot be an observable. Only the combination $ \phi^A \phi_A $ can possibly have meaning. This leads me to believe that one needs a different construction to properly describe the Higgs field in MMSW.\\
	There is one way, later also used for Yang-Mills fields, to do it properly. I constructed an action by means of an auxiliary field  which, on-shell, reproduces the action of a scalar field. The scalar $ \phi $ itself is without any indices with respect to $ SO(2,3) $, so it transforms trivially, but the auxiliary $ V^A $ is a \textbf{5}-vector zero-form whose covariant derivative I suppose to be $ DV^A = dV^A + \frac{1}{\rho}\theta^a\delta^A_5 V_a - \frac{1}{\rho}\theta^a\delta^A_a V_5 $, so without spin connection part. The action is as follows:
	\begin{align}\label{4-26}
		S &= \frac{-1}{3!\rho}\int_M (D\phi + \frac{1}{2} V_A D\tau^A)\wedge V^B (\circledast D\tau^3 )_B\\
		&\to \int_M (D_k\phi - \frac{1}{2} V_k )V^k vol
	\end{align}
	where, in the standard gauge, the equation of motion for $ V $ gives $ V_A D\tau^A = D\phi $ or $ V_k = D_k\phi $. When reinserted into the action, this reproduces the familiar action for a massless scalar field. In addition, the scalar is a viable observable here and the equations of motion are linear. It seems that the coupling is not exactly correct.\\
	As one can see, there are multiple ways to implement a scalar field. However, both seem unnatural and require the use of auxiliary fields, while one even introduces mandatory nonlinearities to the theory. 
	\newpage
\section[Yang-Mills fields]{Yang-Mills fields}\label{Gauge}
	 Yang-Mills(YM) fields are in a sense more and less problematic than scalars. Since they transform nontrivially under diffeomorphisms, one can find more ways to construct actions which might be SO(2,3)-invariant. However, finding an action which exactly reproduces the YM action\footnote{Here, I assume that $ \text{tr}(T^aT^b) = \delta^{ab} $ for a the Lie algebra generators of a gauge group $ N $ in order to make the numerical factors fit for Maxwell theory as well.}  \begin{equation}\label{4-27}
		S_{YM} = \frac{-1}{2g^2}\int_{M} \text{tr}(F\wedge \ast F)
	\end{equation} has so far been unsuccessful. Mainly, the absence of a metric makes a proper Hodge dual impossible. Here, I will give some ideas and a solution of my own as an approach this problem through a first-order formulation of YM theory. There was no kinetic term for these fields available, as Fukuyama's result does not fulfil the criteria put forth in section \ref{PrimCons}.\\
	Let me assume that, for lack of a better theory, there is no unification of gravity and YM theories, so that one does not describe them through one single field. Then, one will optimally use the gauge connections as one-forms $ B_\mu dx^\mu $ just like the MMSW connection form and one will have a single principal bundle from which the total connection stems. However, as seen in section \ref{Unirreps}, one cannot have these connection forms describe particles unless they transform nontrivially under the internal $ SO(2,3) $. So, the minimum one has to describe YM is a \textbf{5}-valued one-form $ B^A_\mu $. On a flat background, this will live in the representation 
	\begin{equation}\label{4-28}
	\left(1,1\right) \oplus \left(1,0\right) \oplus \left(0,1\right) \oplus \left(\frac{1}{2},\frac{1}{2}\right) \oplus \left(0,0\right)
	\end{equation} as described in section \ref{PrimCons}. If one wants to describe spin-1 particles with this gauge potential, one needs to remove the scalar and tensorial part at the very least. To describe YM, one needs to remove the parity-invariant spin-1 part as well. One way would be to only leave the $ A=5 $-part, which would transform properly under Lorentz again and could have the usual gauge constraints imposed on it again. The simplest way to achieve this is to only use the combination 
	\begin{align}\label{4-29}
		\tilde{B}^A &= B^C \tau_C \tau^A\\
		&\to B^5 \delta^A_5
	\end{align}but for now I shall concentrate on finding the most general kinetic term for such a field.\\
	To find such a term, I will assume that the action has an antisymmetric symbol in it, like all other actions considered so far do. This, with the requirement of integrating over a four-form, gives two numerical constraints on the numbers of factors in the kinetic term. Let $ N_{B},N_{DB},N_{\tau},N_{D\tau} $ denote the numbers of $ B,DB, \tau, D\tau $-factors, respectively. Then one has 
	\begin{align}\label{4-30}
		N_{B}+2N_{DB}+N_{D\tau} &= 4\\
		N_{B}+N_{DB}+N_{\tau}+N_{D\tau} &= 5
	\end{align}where $ D_\tau $ is either $ 1 $ or $ 0 $. Subtracting the two gives $ N_{DB} +1 = N_\tau $, which can only be satisfied for $ N_{DB}=0, N_\tau = 1 $. So it is impossible to include a derivative of $ B $ in this structure. One thus needs a different structure for this problem. Another attempt might be an action like
	\begin{align}\label{4-31}
		S = \int_{M} \text{tr}((DB)^A\wedge(DB)^B) \eta_{AB}
	\end{align}In the case that $ B^a = 0 $, this reduces to 
	\begin{align}\label{4-32}
		S &= \int_{M}\text{tr}((DB)^5\wedge(DB)^5 - \frac{1}{\rho^2}\eta_{ab}\theta^a \wedge\theta^b\wedge B^5 \wedge B^5 )\\
		&= \int_{M}\text{tr}(dB^5\wedge dB^5)
	\end{align}which can be identified as the \textit{self-dual} YM action or the \textit{theta term} in nonabelian theories. It is also just the second Chern class of the connection. This is purely topological and does not have any coupling to gravity. A way to circumvent this restriction might be to use a dynamical mechanism of "flattening" the connection to its $ 5 $th component, which could give a small coupling to the MMSW field. As it stands, though, this is unsatisfactory as one cannot guarantee this approach to exactly reproduce the YM action.\\ 
	A different action can give rise to the YM action, however. To my knowledge, this has not been done in the literature before. 
	 \\
	One can get the YM action for a gauge group $ N $ from an $ \epsilon $-type action and an additional, auxiliary field which is a $ \mathbf{10}\otimes \text{ad}(N) $-valued zero-form:
	\begin{align}\label{4-33}
		\begin{split}
		S  &=  \frac{\alpha}{4\rho}\int_{M} \text{tr}_{G}(\text{tr}_{N}(F C)\wedge\circledast(D\tau\wedge D\tau)) \\
		&=  \frac{\alpha}{2\rho}\int_{M} \text{tr}_{N}(F C^{AB})\wedge \circledast(D\tau\wedge D\tau)_{AB} \\
		&\to \alpha \int_M vol \, \text{tr}_{N}(F_{\mu\nu} C^{ab}) e^\mu_ae^\nu_b
		\end{split}
	\end{align}
	after changing to the standard gauge. If one adds a term to induce $ F = D\tau^A \wedge D\tau^B C_{AB} $, this recreates the YM action exactly. The situation is thus similar to the scalar field in that one has to switch to a first-order description and needs auxiliary fields, but distinct in that the auxiliaries do not transform under the same representation as the primary field. In fact, the connection form transforms \textbf{trivially} in this approach, but the dynamics may relate its field strength to the auxiliary $ C $ to give it an interpretation as a particle carrier. In addition, this setup preserves the role of each gauge connection perfectly and one can write a complete action for the entire internal gauge group.\\
	One action which does give rise to the sought-for relation is
	\begin{align}\label{4-34}
	\begin{split}
	S_{C=F} &= \frac{-\alpha}{8\rho}\int_{M} \text{tr}_{G}(\text{tr}_{N}((C_{AB}D\tau^A\wedge D\tau^B) C)\wedge\circledast(D\tau\wedge D\tau))\\
	&= \frac{-\alpha}{4\rho}\int_{M} \text{tr}_{N}((C_{AB}D\tau^A\wedge D\tau^B) C^{RS})\wedge\circledast(D\tau\wedge D\tau)_{RS}\\
	&= \frac{-\alpha}{4\rho}\int_{M} \text{tr}_{N}(C_{AB} C^{RS})D\tau^A\wedge D\tau^B\wedge\circledast(D\tau\wedge D\tau)_{RS}
	\end{split}
	\end{align}
	The equation of motion for $ C $ gives $ \tilde{C}:= C_{AB}D\tau^A\wedge D\tau^B = F $.
	The full action is simply
	\begin{equation}\label{4-35}
		S  =  \frac{\alpha}{4\rho}\int_{M} \text{tr}_{G}(\text{tr}_{N}\left[(F - \frac{1}{2}\tilde{C})C\right]\wedge\circledast(D\tau\wedge D\tau))
	\end{equation}
	and has nontrivial coupling to gravity. This is highly similar to the other actions found here, unlike the Fukuyama scalar action, which does not include $ \tau $ without its derivative. As the gauge potential appears only in combination with $ C $, the auxiliary field must transform like $ F $ does, to make the action gauge invariant.
	It is obvious that only the four-tensor part of the auxiliary field plays a role in the action. As such, its interpretation as a spin-1 particle is a given. In addition, the gauge invariance of the theory gives a clear indicator that this action describes, through convoluted means, a $ D(2,1) $-theory, as these are the only spin-1 representations that admit gauge invariance. This is implemented as \begin{equation}\label{4-36}
		(DC)^{AB} = dC^{AB} + s\frac{1}{2}f_{cd,EF}^{AB} \, \omega^{cd} C^{EF} + \frac{q}{\rho} f_{5b,EF}^{AB} \theta^b C^{EF}
	\end{equation}with $ s=1, q=2 $. Again, one would need to check explicitly that this lives in the supposed unirrep, but this will not be done here as a quantal version of this theory is still missing.
	One can thus find a way to introduce YM fields, but it also requires a somewhat artificial construction. Most importantly, in giving gravity the form of a YM type action, one had to sacrifice the same form for spin-one gauge fields. This clashes heavily with the hope behind MMSW, which is to describe all gauge bosons in an equivalent way. 
	\newpage
\section{Matter-Matter couplings}
	The requirements put forth in section \ref{PrimCons} highly restrict possible renormalisable matter-matter couplings, which warrants a separate discussion. Derivative couplings will not be dealt with here, as they have similar complications as the actions given above. Also, some derivative couplings can be naturally incorporated within terms that, in the unbroken phase, do not include derivatives. In addition, the usual couplings from YM theory generalise readily and do not need adjustment, as they are completely encoded in covariant derivatives. The results of this section are not based on previous work by other authors.\\
\subsection{Self-interaction}\label{SelfInt}
	First, one might be interested in self-coupling. As is well-known, there are no renormalisable spinor self-interactions on 3+1D Minkowski space, as simple dimensional analysis will tell. The situation in MMSW is similar: As Dirac spinors can only appear in pairs in the action, one cannot use terms which include more than two Dirac fields without creating a non-renormalisable interaction.
	For scalars, which have mass dimension one, I constructed the usual cubic interaction vertex, while the quartic one was given by Fukuyama\cite{Fukuyama}. However, the embedding of scalar fields into \textbf{5}-vectors makes this more challenging, as manifest $ SO(2,3) $-invariance must be preserved. One can use the first generic coupling for the quartic coupling:
	\begin{align}\label{4-37}
		S_4 &= \frac{\lambda_4}{\rho} \int_{M} \left(\phi^A\phi_A\right)^2 \, \circledast(D\tau\wedge D\tau\wedge D\tau\wedge D\tau)\\
		&\to \lambda_4 \int_{M} vol \,  \left(\phi^a\phi_a\right)^2 + (\phi^5)^4 \, 		
	\end{align}
	A similar form was given by Fukuyama as well.
	And similarly, for the second generic coupling using a \textbf{5}-vector for the cubic coupling:
	\begin{align}\label{4-38}
		S_3 &= \frac{\lambda_3}{4! \rho} \int_M \phi_M\phi^M \phi^A D\tau^B\wedge D\tau^C\wedge D\tau^D\wedge D\tau^E \epsilon_{ABCDE} \\
		&\to \frac{\lambda_3}{\rho} \int_M vol \,  \left( \phi^5\phi_a\phi^a + (\phi^5)^3\right)
	\end{align}
	One can recognise the cubic and quartic terms easily. However, the $ \phi^a $-components also make an appearance. One cant simply use the formula (\ref{4-22}) to eliminate these now, as the interaction term will modify the relation. Indeed, that elimination process will inevitably give dependencies on the coupling constants $ {[\lambda_4]} = 0 = {[\lambda_3]}  $ \footnote{The cubic coupling has been changed arbitrarily to be dimensionless in favour of a $ \frac{1}{\rho} $-prefactor, which has interpretive advantages in the full theory. One may of course use a more general, dimensionful constant, but this just amounts to a rescaling. See the discussion in section \ref{Discussion}.} and induce additional, higher order nonlinearities when reinserted into the action. As such, one can see that it is extremely difficult to introduce free scalars into this picture. This might not be of high importance, however, as only one specialised scalar field is known to be fundamental, and it transforms nontrivially under additional internal symmetries as well. The nonlinearities shown here could perhaps even give rise to more complicated potentials, giving the Higgs a richer vacuum structure. However, the interactions might also spoil the invertibility for $ \phi_a $ and thus make it difficult to determine whether the field still describes a spin-0 particle.
	As it is the more realistic case, consider the quartic theory, \begin{align}\label{4-39}
		S &=  \frac{1}{\rho}\int_M \epsilon_{ABCDE} \; \left[\frac{1}{3!}\phi^A D\phi^B + \frac{\lambda_4}{4!} \left(\phi^A\phi_A\right)^2 \tau^A D\tau^B\right] \;\wedge D\tau^C\wedge D\tau^D\wedge D\tau^E \\
		&\to \int_{M} vol \, \frac{1}{\rho}\phi^A M_{AB}^\mu (\partial_\mu \phi^B + \mathcal{A}^B_C \phi^c) + \lambda_4  \left(\phi^A\phi_A\right)^2 
	\end{align} 
	which give the EoM
	\begin{align}\label{4-40}
		(D_{\mu}( \phi^A M_{AB}^\mu)) - M_{BA}^\mu (D_\mu \phi)^A = 4 \rho \lambda_4  \phi_B \left(\phi^A\phi_A\right)
	\end{align} In the case of $ D_\mu (e^\mu)_a = 0 $, one gets the relation \begin{equation}\label{4-41}
		\phi_a = -2\rho \partial_a \phi^5 - 4\rho^2 \lambda_4 \phi_a \left(\phi^b\phi_b + (\phi^5)^2\right)
	\end{equation} which is not invertible for $ \phi_a $. One can, of course, still try to use perturbation theory in $ \lambda_4 $, though it will necessarily make more and more interaction terms appear in higher orders. This is clear as one has integrated out the four-vector components of the field. For example, the solution to first order in $ \lambda_4 $ already gives 
	\begin{align}\label{4-42}
		\phi_a &= \phi^{(0)}_a + \lambda_4 \phi^{(1)}_a\\
		&=-2\rho \partial_a \phi^5\left[1 + 4\rho^2\lambda_4 \partial_a \phi^5 (4\rho^2 \partial_b \phi^5\partial^b\phi^5 + \phi^5 \phi^5)		\right] 
	\end{align} and so a perturbation series for quantities involving only $ \phi^5 $ will become extremely difficult for strong couplings. Overall, one can see that scalar fields are a weak point of MMSW, unless those difficulties can be avoided by considering the full \textbf{5}-vector.\\
	A last fact about special solutions of the full, unbroken equations of motion is noteworthy: When one considers $ \phi^a = 0 $-solutions, one has 
	\begin{align}\label{4-43}
	& 2e^\mu_a\partial_{\mu}\phi^5 + \phi^5 (D_\mu\phi^\mu)_a = 0& \lambda_4 (\phi^5)^3 + \frac{2}{\rho^2}\phi^5 = 0&
	\end{align}which implies that one can, for negative $ \lambda_4 $, find constant solutions which are nonzero: $ \text{VEV} = \pm\sqrt{\frac{-2}{\rho^2 \lambda_4}} $. Thus one can, in principle, achieve a Higgs-type mechanism for scalar fields even in the full theory, though the resulting dynamics may differ wildly.
	\newpage
\subsection{Yukawa interactions}\label{Yukawa}
	Yukawa-type couplings can be easily incorporated into the MMSW formalism. I took either of the generic coupling actions given in \ref{PrimCons} to construct an interaction term for a real scalar and a Dirac spinor in multiple ways, with appropriate coupling constants:
	\begin{enumerate}\label{4-44}
		\item $ S_1 = \frac{1}{4!\rho}\int_{M} \bar{\psi}\gamma^A\psi \phi_A \, \circledast(D\tau\wedge D\tau\wedge D\tau\wedge D\tau) $
		\item $ S_2 = \frac{1}{4!\rho}\int_{M} \bar{\psi}\gamma^A\gamma^5\psi \phi_A \, \circledast(D\tau\wedge D\tau\wedge D\tau\wedge D\tau) $
		\item $ S_3 = \frac{1}{4!} \int_M \bar{\psi}\Sigma^{AM}\psi \phi_M D\tau^B\wedge D\tau^C\wedge D\tau^D\wedge D\tau^E \epsilon_{ABCDE}$
		\item $ S_4 = \frac{1}{4!} \int_M \bar{\psi}\Sigma^{AM}\gamma^5\psi \phi_M D\tau^B\wedge D\tau^C\wedge D\tau^D\wedge D\tau^E \epsilon_{ABCDE}$
		\item $ S_5 = \frac{1}{4!} \int_M \bar{\psi}\phi^A\gamma^5\psi  D\tau^B\wedge D\tau^C\wedge D\tau^D\wedge D\tau^E \epsilon_{ABCDE}$
		\item $ S_6 = \frac{1}{4!} \int_M \bar{\psi}\phi^A\psi  D\tau^B\wedge D\tau^C\wedge D\tau^D\wedge D\tau^E \epsilon_{ABCDE}$
	\end{enumerate}
	These give various derivative and non-derivative spinor-scalar couplings, of which only $ S_5,S_6 $ give the regular Yukawa couplings freely. $ S_3,S_4 $ both give nonlinear interactions involving derivatives, while $ S_1,S_2 $ give a combination of both. Thus, all couplings present in the standard model can be incorporated into the MMSW formalism. Obviously, this will modify the relation (\ref{4-22}) for the four-vector part as well and create even more effective couplings between the scalar and the spinor. 
	
\section{The role of Torsion}\label{Torsion}
	It is worth clearing up a few issues regarding the presence of nonzero torsion in the dynamics. As was put forth elsewhere \cite{VasilievTheory}, Diffeomorphisms and internal transformations are part of a semidirect product group. The two do not commute and there is a special relationship between them when the field strengths $ F $ of the gauge connection vanish. The change of a connection-one form under internal transformations with parameter $ \epsilon $ and diffeomorphisms with parameter $ \xi $ is 
	\begin{equation}\label{4-45}
		\delta_{\xi,\epsilon}(A) = d\epsilon + \mathcal{L}_\xi A
	\end{equation}
	Under certain redefinitions of fields, one can write it as 
	\begin{equation}\label{4-46}
		\delta_{\xi,\epsilon}(A) = d\epsilon + i_{\xi}F
	\end{equation}
	If the field strength $ F $ vanishes, one may do even more, as one always has 
	\begin{equation}\label{4-47}
		\mathcal{L}_\xi A = D(i_{\xi} A) + i_{\xi}F
	\end{equation}just by a generalisation of Cartan's magic formula. When the field strength vanishes, one can identify the changes on the left with the changes on the right, or diffeomorphisms with gauge transformations. For a gauge algebra which has dimension lower than $ \text{dim}(M) = 4 $, one does not have enough free parameters to do so. However, when one has at least dimension $ 4 $, one can set a part of the curvature to zero to have exact correspondence between the two transformations. In Einstein gravity, one sets the torsion to zero, which gives exact correspondence between local translations and diffeomorphisms and reduces the total gauge degrees of freedom from $ 10 $ to $ 6 $. These $ 4+4+6 = 14 $ constraints and transformations then allow one to reduce a general, non-symmetric metric to the two propagating degrees of freedom of a massless graviton.\\
	Now, this would be nice and simple, but the dynamics put constraints on how much this is possible. To understand this, one has to view MMSW as an extension of EG in the sense of Einstein-Cartan theory\cite{CartanReview}. Torsion is allowed to be, in principle, nonzero, by allowing for more general metric connections. In Einstein-Cartan theory, one just adds a contorsion tensor to the Levi-Civita connection which results in an additional equation of motion for the torsion tensor. This equation is algebraic and relates the spin and torsion tensors. This same effect happens in the Palatini formalism when incorporating spinors. This makes the question whether one can feasibly set the torsion to zero nontrivial. To investigate this, a simple case such as a spinor coupled to MMSW should suffice.\\
	The action under consideration is 
	\begin{equation}\label{4-48}
		S = \frac{1}{4\alpha\rho}\int_{M} F^{AB}\wedge(\circledast F)_{AB} + \frac{1}{3!\rho}\int_{M} \bar{\psi} (\circledast D\tau^3)_A \gamma^A\wedge iD\psi
	\end{equation}
	with equations of motion w.r.t $ \mathcal{A} $:
	\begin{equation}\label{4-49}
	\frac{2}{4\alpha \rho} (D\circledast F)_{AB} = \frac{i}{3!\rho} \bar{\psi}\gamma^C\Sigma_{AB}\psi (\circledast D\tau^3)_C  + \frac{i}{\rho} \bar{\psi}\gamma^C\psi \tau_B (\circledast D\tau^2)_{AC} 
	\end{equation}
	which, in the standard gauge, are
	\begin{align}\label{4-50}
		\frac{1}{2}\epsilon_{abcd} \theta^b\wedge F^{cd} &= \frac{i\kappa }{6\rho} \bar{\psi}\gamma^c\gamma_a\psi \epsilon_{cmnk} \theta^m\wedge\theta^n\wedge\theta^k -i\kappa \bar{\psi}\gamma^b\psi \epsilon_{abcd} \theta^c\wedge\theta^d \\
		(D\circledast F)_{ab} &= -\frac{i\kappa}{3\rho} \bar{\psi}\gamma^c\Sigma_{ab}\psi \epsilon_{cmnk}\theta^m\wedge\theta^n\wedge\theta^k
	\end{align}
	The first equation reduces to Einstein's field equation. The second one can be written as 
	\begin{equation}\label{4-51}
		dF^{rs} = \frac{-i\kappa}{3! \rho^2} \epsilon_{cmnk} \epsilon^{rsab} \bar{\psi}\gamma^c\Sigma_{ab}\psi \theta^m\wedge\theta^n\wedge\theta^k
	\end{equation}
	which shows that the exactness of the Lorentz part of the curvature is broken by the presence of matter with spin. Using $ \gamma^c\Sigma_{ab} = \frac{-i}{2}(\gamma_{a}\delta^c_b - \gamma_{b}\delta^c_a) $, which one finds through commutators, 
	\begin{align}\label{4-52}
		\begin{split}
		dF^{rs} &= -\frac{\kappa}{3! \rho^2} \bar{\psi}\gamma_a\psi\, \theta^m\wedge\theta^n\wedge\theta^k\epsilon_{bmnk}\epsilon^{brsa}\\
		&= -\frac{\kappa}{3! \rho^2} \bar{\psi}\gamma_a\psi\, \theta^r\wedge\theta^s\wedge\theta^a\\
		&= -\theta^r\wedge\theta^s\wedge\,  \frac{\kappa}{3! \rho^2}\bar{\psi}\gamma_a\psi\theta^a
		\end{split}
	\end{align}
	so that clearly, the torsion tensor appears at no point of the calculation. However, as mentioned before, there is manifestly nonzero torsion in the Palatini variation. By using the relation of the spin tensor to the torsion, one actually has
	\begin{align}\label{4-53}
		\epsilon_{abcd} \theta^c\wedge T^d + \frac{\kappa}{3!} \epsilon_{mnkl} \bar{\psi}(\gamma_{a}\delta^m_b - \gamma_{b}\delta^m_a)\psi \theta^n\wedge\theta^k\wedge\theta^l = 0
	\end{align}or
	\begin{align}\label{4-54}
	4\theta^u\wedge T^v + \frac{\kappa}{3} \delta^{uva}_{nkl} \bar{\psi}\gamma_{a}\psi \theta^n\wedge\theta^k\wedge\theta^l &= 0 \\
	4\theta^u\wedge T^v + 2\kappa  \bar{\psi}\gamma_{a}\psi \theta^u\wedge\theta^v\wedge\theta^a &=0
	\end{align} and so finally
	\begin{align}\label{4-55}
		T^a = \frac{\kappa}{2} \bar{\psi}\gamma_b\psi \theta^b \wedge \theta^a = \frac{\kappa}{2} J_b \theta^b \wedge \theta^a
	\end{align}
	If one now sees this as a relation for $ \omega $, by the definition $ \dot{D}\theta = T $, one has
	\begin{align}\label{4-56}
		d\theta^a + \omega^{ab}\wedge\theta_b = \frac{\kappa}{2} J_b \theta^b \wedge \theta^a
	\end{align}and so, by using\cite{VasilievTheory}
	\begin{align}\label{4-57}
		T_{abc} &= Y_{abc} + \omega_{acb} - \omega_{abc} \\
		Y^a_{\mu\nu} &:= (d\theta^a)_{\mu\nu}
	\end{align}where indices are raised and lowered by vierbeins, one achieves by adding cyclic permutations:
	\begin{align}\label{4-58}
		\omega^{ab}_\mu &= \frac{1}{2}\theta^d_\mu \eta_{cd} \left(Y^{abc} + Y^{bca} - Y^{cab}\right) - \frac{1}{2}\theta^c_\mu \eta_{cd} \left(T^{abc} + T^{bca} - T^{cab}\right)\\
		\omega^{ab}_\mu &= \omega^{ab}_{(0)\mu} - \frac{1}{2}\theta^d_\mu \eta_{cd} \left(T^{abc} + T^{bca} - T^{cab}\right)
	\end{align}	where $ \omega_{(0)} $ denotes the torsion-free part of the connection.\\
	In this case, $ T^{abc} = \frac{\kappa}{2} \left(\eta^{ca}J^b-\eta^{cb}J^a\right) $ and so
	\begin{align}\label{4-59}
		\omega^{ab}_\mu &= \omega^{ab}_{(0)\mu} - \frac{\kappa}{2} \left( \theta^a_\mu J^b - \theta^b_\mu J^a\right)\\
		\omega &=\omega_{(0)} + K
	\end{align}This is manifestly dependent on the matter flux $ J^a = \bar{\psi}\gamma^a\psi $. Now that the equation of motion has been solved for the spin connection, the torsion has been completely eliminated from the dynamics. It is not constrained to vanish, but instead the spin connection now is a function of the vierbein and matter flux in such a way that it incorporates the dynamics of the torsion. One can now, in principle, reinsert this expression into the action. When defining $  R_0 = d\omega_{(0)} + \frac{1}{2}{[\omega_{(0)},\omega_{(0)}]}$, this will lead to the replacements
	\begin{align}\label{4-60}
		&R\to R_0 + \dot{D}_0(K)& D\psi \to D_0\psi + K\psi&
	\end{align}which will give new terms in the action, now only dependent on $ \theta $. In particular, the Dirac action in the standard gauge will have an additional term
	\begin{align}\label{4-61}
		\frac{3}{4}\kappa \bar{\psi}\gamma^a\psi \, \bar{\psi}\gamma_a\psi
	\end{align}
	which is nonrenormalisable in 4D. Similarly, the gravity action generates new terms like $ 2\dot{D}_0K\wedge \theta\wedge \theta + \frac{1}{2}{[K,K]}\wedge \theta\wedge \theta $.\\
	Now the question how the non zero torsion modifies the gauge degrees of freedom is more easy to answer. In general, the torsional part of the field strength, as used in (\ref{4-47}), is nonzero on-shell. Of course, one can replace it and the spin connection in the action by appropriate expressions, which gives rise to effective theories whose torsion is non-manifestly nonzero. In doing so, however, the effective EM tensor becomes the Hilbert one and the coupling prescription is fixed at the cost of new terms which are quadratic in the matter fields and may couple to the Riemann curvature directly. \\
	This, however, masks the fact that the most natural coupling description for spinors makes the gauge structure different. One can no longer represent diffeomorphisms as gauge transformations of the field. The invariance now should consist of the diffeomorphism and the gauge group seperately. Since in the standard gauge, one had to impose four constraints for each diffeomorphism to ensure that physical components do not mix with unphysical ones\cite{KucharDiffs}, one could reduce the 16 degrees of freedom of the vierbein to the two of the graviton. Now, it would seem that one has four more gauge freedoms, which would be more than enough to eliminate the gauge field completely. However, since the field strength and even torsion do not in general vanish, there seem to be fewer constraints one can impose, exactly those four one previously had when local translations corresponded to diffeomorphisms. Then, the torsional and Riemann parts were seperate sectors and were required to mix so that the propagating degrees of freedom stayed the same. This was implemented as four constraints on the gauge field connecting diffeomorphisms and local translations. Now, this is no longer necessary and one does not have these constraints anymore in favour of proper local translation invariance.\\
	One can thus see that there are still only two propagating degrees of freedom even when the torsion is nonzero, at least after eliminating it and the free spin connection from the action. The gauge group takes on a role more closely to other gauge theories and diffeomorphisms have to be treated in MMSW as in Yang-Mills theory.

\section{When the vierbein vanishes}\label{meme}
	In this section, I want to note a peculiar feature of this formalism. When one formulates the actions manifestly $ SO(2,3) $-invariantly, one finds that one can have $ \mathcal{A}=0 $ without the actions vanishing. This is quite different from GR, where the couplings were such that the vanishing of the vierbein would eliminate all coupled matter actions. Previously, these cases had been discarded as unphysical, as there is no way to interpret these configurations of $ \mathcal{A} $ or $ \theta $ as spacetime metrics. However, in light of the stability of the actions given here even when the MMSW field vanishes, I plead for inclusion of this case as a legitimate one. As an example, I will consider a single spinor field, which reads
	\begin{equation}\label{4-62}
		S = \frac{1}{3! \rho} \int_M \bar{\psi} (\circledast d\tau^3)_A \gamma^A\wedge d\psi
	\end{equation}and where all instances of the gravitational field have been removed. It seems that one could make this vanish, too, by going to the standard gauge or any other constant section $ \tau $. However, one has to remember that to the zero-field configuration, all pure gauge configurations of $ \mathcal{A} $ are connected. As such, choosing it to be zero amounts to choosing a gauge up to global transformations. So, one cannot reduce $ \tau $ to the standard form anymore, unless it is already in a globally equivalent form. As such, the action in question need not necessarily vanish and one can even consider its equations of motion, which turn out to be, when varying wrt. $ \bar{\psi} $ and $ \tau^A $, respectively, by defining $ Q_A := \,\epsilon_{ABCDE}d\tau^B\wedge d\tau^C\wedge d\tau^D \gamma^E  $:
	\begin{align}\label{4-63}
		& \gamma^A(\circledast d\tau^3)_A \wedge d\psi = 0  & \bar{\psi}Q_A \wedge d\psi = 0 &
	\end{align}where the boundary term for the second equation vanishes identically. These are 8 equations and do not fix $ \tau $ by much, so it seems that one has to provide the tangency map even in the case of vanishing potential. However, there is a natural choice, since for vanishing curvature one has also the gauge where the potential is the one describing Anti-deSitter space -  namely, $ \tau $ sends a point of the base manifold to an equivalent point in the local copy. So, by finding the transformation that reduces the Anti-deSitter potential to zero, one can just as well find an adequate tangency map by applying the same transformation to the natural one for Anti-deSitter. Then, one could easily find insert that expression into the equations of motion and ask for solutions, which seem to have no large obstructions from existing. Indeed, on a compact manifold, constant solutions for $ \psi $ trivially solve the equations, so the relevant question is actually about the form of nonconstant solutions.
	The actual form of these solutions, however, is not the point here. Instead, it is to point out that one can have dynamics, however restrained, even in the absence of a gravitational field. This supports the view that spacetime is an emergent phenomenon in some sense, even though this conceptually very different from modern approaches which seek to construct the spacetime manifold as an effective object. Here, a metric, and thus an interpretation of the gravitational potential in terms of Riemannian geometry is possible through the framework of Cartan geometry, and no effective treatment of the fields is used. One still has the entirety of field theory available, though the theories are all topological in nature. One might see this interpretation of zero MMSW connections as the "simplest" case of spacetime emergence, which is made possible in all cases where the vierbein matrix is nondegenerate, which is almost always the case. Of course, this is only an emergence of the metric as the underlying spacetime manifold is entirely unaffected by all of this.

	\chapter[Discussion]{Discussion}\label{Discussion}
	In this thesis, some aspects of MMSW gravity with respect to coupling were studied, in particular the form of invariant actions and their couplings.
	First, the groundwork of the geometrical background was laid out to explain the presence and significance of the \textit{tangency point map} $ \tau $. The reduction to Einstein gravity was also highlighted. Afterwards, the action and equations of motion of MMSW gravity were studied in both the unbroken and broken phase. Following on that, the representations by which field theories could be classfied in this formalism were presented to lay the groundwork on which the later sections were understood.
	From there on, kinetic terms for spinor and scalar fields were discussed. The action for the scalar field was found to have several issues, among them that Fukuyama's action did not allow for the scalar to be an observable. In addition, it was plagued by nonlinearities. I proposed another action, but its coupling to gravity seems to be incorrect. As such, the problem of the scalar field coupling remains open, though more possibilities have been exhausted.
	
	A similar problem, though with a more clear solution, has been found for Yang-Mills fields. Here, I presented an option of using the gauge field as-is that produces the expected coupling on-shell by means of an auxiliary zero-form. This, together with the spinor action of Fukuyama in a modified form, gives the framework one needs to describe electrodynamics in MMSW as the QED coupling needs no further adjustment.
	Furthermore, some self-interactions and Yukawa couplings were presented. It was found that the Fukuyama scalar is highly difficult to deal with in perturbation theory even at low orders. 
	Following up, an explicit example showed the presence of nonzero on-shell torsion in the dynamics and how to eliminate it. The gauge structure of MMSW in the presence of such unconventional field strengths was explained through a lack of imposable constraints.
	Finally, a bit of light was shone on seemingly pathological configurations of MMSW. These were legitimised through an argument about consistency of matter actions even in the pathological situation.

	The main virtue of MMSW gravity is its clear similarity to Yang-Mills field theories: The gravitational field is integrated into a gauge connection one-form, which puts them on more even footing. In addition, it reduces the number of free dimensionful parameters of the theory - instead of the cosmological and Newton's constant, one parametrises the theory by a tiny dimensionless coupling and a fundamental length scale $ \rho $. The theory predicts a coupling to the Euler characteristic of the manifold, which is a quadratic gravity term that does not affect the equations of motion. This is in principle only testable through global means, which are not available to my knowledge, or by looking for quantal effects of MMSW that depend on the Euler characteristic. The latter seems untestable as of now, too, but in principle one should be able to find the value of these constants given better measurements.
	
	Most predictions are indeed the same as for regular gravity: Spinor couplings are equivalent to the usual ones, just as for gauge fields. Scalar fields have been shown to be more problematic and will need more attention in the future, but a more elaborate setup might remedy these issues. In the presence of matter with spin, there is nonzero torsion, which is similar to Einstein-Cartan or Palatini gravity. This has nontrivial effects on, for example, fermion shape and thus the UV cutoffs of the theory\cite{TorsionDirac}.
	
	There is, however, a large problem with these couplings and actions, nonetheless. To embed other bosons into this formalism in a manifest way, one has to give up the usual form of the actions. In particular, $ s=1 $ gauge fields are no longer described by a Yang Mills-type action. This defeats the purpose of the formalism, as it was intended to describe gravity through a gauge connection and thus make it similar to a Yang Mills theory. One can learn from this, however, that gravity is indeed very different from Yang Mills theories.
	
	Now, given that the effective cosmological constant we measure in $ \Lambda CDM $ is positive instead of negative, one cannot apply many of the formulas used here to estimate the true value of $ \rho $. Since the $ q $-values for the masses of fermions are extremely large, one will not be able to tell its value even if one had knowledge of the lightest fermion's mass, for example of the neutrinos. Instead, one might take a different approach to find it:
	
	If one takes a look at most, if not all of the $ SO(2,3) $-invariant actions in this thesis, one will quickly see that they all share a common factor of $ \frac{1}{\rho} $. If one applied standard quantum field theory rules, one could interpret the kinetic terms as those of effective theories with coupling dependency $ \frac{E_\rho}{E} $, so associated with an infrared cutoff of value $ E_\rho = \frac{1}{\rho} $. This also implies an upper length scale to which the theory can be considered valid. Since general relativity has been tested to length scales of the size of the universe, one can guess that in this picture, the value of $ \rho $ must be at the very least larger than the radius of the observable universe, if not larger. However, this analogy is to be taken with caution as the theory here is not a usual quantum field theory, but a topological one. Also, if it were still a good approach, one would inevitably run into issues with quantisation as the cutoff implies the absence of the usual trivial infrared fixed point. In any case, the global factor of $ \frac{1}{\rho} $ in all actions does strike my interest. Only a fully quantum analysis will be able to answer all of the questions mentioned thus far.
	
	Overall, many interesting venues of study remain open in this topic. The now available actions for gauge fields need to be analysed further to ensure they do give rise to the same quantum theory. More generally, quantum versions of these topological theories must be found. Additionally, it would be interesting to study the role of the singleton representations in this theory and their relation to gravity, as both kinds of fields have to couple universally to others. This would need a description in terms of the conformal boundary of the spacetime, thus delving even further into the AdS-CFT correspondence which one suspects to play a role in a theory with internal AdS group. This goes back to Fukuyama, who derived the $ SO(2,3) $-symmetry from a conformal model which breaks down to an AdS-invariant system. Finally, the similarities between Higgs-Chern-Simons forms and MMSW are intriguing and suggest further connections between gravity and topological theories.
	
	The role of MMSW is a connection between the traditional world of local metric theories of gravitation and the emergent area of topological gravity. It does not, however, give the gauge theoretic description of gravity that it seems to provide at first glance. Its importance of MMSW is thusthat of an interesting reformulation which, however, comes with many problems.
	
	In my opinion, the theory has many features which are at least equally balanced out by striking issues. The ease with which it can be extended from EG, its superficial removal of UV divergences and natural mass terms initially give an impression of a promising concept which is held back by its complications with coupling. If these problems can be overcome in a satisfactory way, then one should pick up studies in this topic again. 
	
	\chapter{Appendix}

\section{Conventions}
	Here, I will declare some of the conventions I use in the thesis. Throughout it, I use $ a,b,c,d,e $, or lower Latin indices, for component expressions of the algebra $ \mathfrak{so}(1,3) $, running from $ 0 $ to $ 3 $, and similarly, capital Latin indices $ A,B,C,D,E,\dots $ for ones of $ \mathfrak{so}(2,3) $.These run in $ 0,1,2,3,5 $, as is conventional. In addition, I work in four-dimensional spacetime manifolds, whose coordinates I index with lower greek glyphs $ \alpha \beta \gamma \dots $ which run from $ 0 $ to $ 3 $.\\
	I use natural units, so that only mass dimensions are given. I assign, for differential forms, $ {[dx^\mu]}=-1 $ and $ {[\frac{\partial}{\partial x^\mu}]} = +1 $, so that taking an exterior (covariant) derivative results in no change of mass dimension -  $ {[d]} = {[D]} = 0 $.
	I also often use the notation $ \dot{D} $ to denote the covariant derivative with respect to a Lorentz connection.

\subsection{Lie algebras}\label{LAConv}
	I often work with multiple indefinite orthogonal groups in this thesis.\cite{Weyl1997,Hamermesh2} Given a symmetric bilinear form of signature $ (p,n-p) $ on an $ n $-dimensional real vector space, most often in terms of a matrix with respect to some fixed basis, $ SO(p,n-p) $ is the group of matrices that leave the bilinear form, or its matrix, invariant.
	The most important case here is $ n=5,p=2 $, where the bilinear form's matrix is $ \eta_{AB} = \text{diag}[1,-1,-1,-1,1] $.
	The group matrices $ X $ thus satisfy the defining relation
	\begin{align}\label{6-1}
		X^A_C \eta^{CD} X^B_D = \eta^{AB}
	\end{align}
	The group $ SO(2,3) $ is a \textit{Lie group} - it is a differentiable manifold in addition to being a group. It thus admits the well-known Lie correspondence to its Lie algebra $ \mathfrak{so}(2,3) $, defined as the set of left-group-action-invariant vector fields and identified with the tangent space of $ SO(2,3) $ at the identity.
	I use the standard basis and its commutation relations:\cite{Schwartz}
	\begin{equation}\label{6-2}
	{[M_{AB},M_{CD}]} = (\eta_{AC} M_{BD} + \eta_{BD} M_{AC} - \eta_{AD} M_{BC} - \eta_{BC} M_{AD})
	\end{equation} 
	and similarly for explicit representations of those generators but with a $ -i $ in front of the braces, by assigning $ \pi(M_{AB})= T_{AB} = i t_{ab} $, where, $ t_{ab} $ are the physics conventional generators. This gives for the structure constants
	\begin{equation}\label{6-3}
		f^{AB,CD}_{KL} = \eta^{BC} \delta^A_K \delta^D_L + \eta^{AD} \delta^B_K \delta^C_L - \eta^{BD} \delta^A_K \delta^C_L - \eta^{AC} \delta^B_K \delta^D_L
	\end{equation}
	In particular, note that if $ A \in \{0,1,2,3,5\} $, so the AdS algebra, one has 
	\begin{align}\label{6-4}
		{[M_{5a},M_{5d}]} &= (\eta_{55} M_{ad} + \eta_{ad} M_{55} - \eta_{5d} M_{a5} - \eta_{a5} M_{5d})\\
		  &= \eta_{55} M_{ad}
	\end{align}
	
	I use the commutator of Lie algebra-valued forms often. It is defined with respect to a general Lie algebra with generators $ T_a $, for a $ p $-form $ \omega = T_a\otimes\omega^{a} $ and a $ q $-form $ \eta = T_a\otimes\eta^{a} $, as\cite{Nak}
	\begin{align}\label{6-5}
	{[\omega,\eta]} &= \omega\wedge\eta - (-1)^{pq} \eta\wedge\omega\\
	&= {[T_a,T_b]}\otimes \omega^a\wedge\eta^b  = f_{ab}^c\, T_c \otimes \omega^a\wedge\eta^b
	\end{align}
	One can use this in the covariant derivative of a Lie algebra-valued form $ \omega $, defined as
	\begin{align}\label{6-6}
		D\omega &= d\omega + {[\mathcal{A},\omega]}\\
		&= T_a \otimes (d\omega^a + f_{bc}^a \, \mathcal{A}^b\wedge\omega^c)
	\end{align}
	With this, one can use an equivalent of the Stokes theorem for this covariant derivative:\cite{VasilievTheory}(See, in particular, a footnote on p.25 of the reference)
	\begin{align}\label{6-7}
		\int D(\omega^{AB}\wedge(\circledast \eta)_{AB}) &= \int d(\omega^{AB}\wedge(\circledast \eta)_{AB})\\
		=\int (D\omega)^{AB}\wedge(\circledast \eta)_{AB} &+ (-1)^p \int \omega^{AB}\wedge (D\circledast \eta)_{AB}
	\end{align}

	For the Killing form of $ SO(2,3) $, I use 
	\begin{equation}\label{6-8}
		\mathcal{K}_{AB,CD} = \frac{-1}{6}f_{AB,GH}^{EF} f_{CD,EF}^{GH} = (\eta_{AC}\eta_{BD} - \eta_{AD} \eta_{BC})
	\end{equation}
	so that, in the fundamental representation, where I use the generators
	\begin{equation}\label{6-9}
		(M_{AB})_{CD} = \mathcal{K}_{AB,CD}
	\end{equation} 
	so that
	\begin{equation}\label{6-10}
		\text{tr}(M_{AB}M_{CD}) = -2\mathcal{K}_{AB,CD}
	\end{equation}
	and 
	\begin{equation}\label{6-11}
		\text{tr}(\Sigma_{AB}\Sigma_{CD}) = \mathcal{K}_{AB,CD} 
	\end{equation} in the physics convention for the \textbf{4}-irrep, defined below.
\subsection{Clifford algebra}\label{CAConv}	
	In this thesis, I use only the complexified Clifford algebra $ Cl_{1,3}(\mathbb{R})_{\mathbb{C}} $, for which I get identities from Schwartz\cite{Schwartz}.
	It is defined through the generating elements $ \gamma^a $, which satisfy 
	\begin{align}\label{6-12}
		\{\gamma^a,\gamma^b\} = 2\eta^{ab}
	\end{align}
	with $\eta^{ab} = \text{diag}[1,-1,-1,-1]$ and the unit $ 4\times4 $ matrix suppressed. I also use 
	\begin{equation}\label{6-13}
	\gamma^5 := -i\gamma^0\gamma^1\gamma^2\gamma^3
	\end{equation}which anticommutes with all the other generators:
	\begin{equation}\label{6-14}
		\{\gamma^\mu,\gamma^5\} = 0
	\end{equation}
	When talking of $ \gamma^A $, I will refer to the set $ \{\gamma^\mu, \gamma^5\} $.

	As a representation of the generators of $ \mathfrak{so}(1,3) $, I use
	\begin{equation}\label{6-15}
		\Sigma^{ab} = \frac{i}{4}  {[\gamma^a,\gamma^b]}
	\end{equation}
	
	I use the Weyl representation of this algebra unless mentioned otherwise. In this basis, one has 
	\begin{align}\label{6-16}
		\gamma^0 (\gamma^b)^\dagger \gamma^0 = \gamma^b
	\end{align} and similarly
	\begin{align}\label{6-17}
	\gamma^0 (\Sigma^{ab})^\dagger \gamma^0 = \Sigma^{ab}
	\end{align}
	
	Using these properties, the objects introduced here can be used for a representation of the Lie algebra of $ SP(4,\mathbb{R}) $, the double cover of the AdS isometry group:
	\\
	Set $ \Pi^a := \frac{\gamma^a}{2} $. Then
	\begin{equation}\label{6-18}
		{[\Pi^a,\Pi^b]} = \frac{1}{4}{[\gamma^a,\gamma^b]} = -i\Sigma^{ab}.
	\end{equation} Thus, the collection $ \Sigma^{AB}= (\Sigma^{ab},\Pi^a) $ give a representation in the style of $ so(2,3) $ by assigning $ \Sigma^{5a} = \Pi^a $. Note that $ \Sigma $ is in the physics convention.
	
	I sometimes use Feynman slash notation:
	\begin{equation}\label{6-19}
	\slashed{V} := V^A \gamma_A
	\end{equation}
	which has nice geometrical properties, both under traces and as-is. For example:
	\begin{align}\label{6-20}
	\slashed{V}\slashed{W} &= \eta(V,W) -2i \Sigma_{AB}V^AW^B.\\
	\slashed{V}\slashed{V} &= \eta(V,W) = V^AV_A.
	\end{align}

	I also define the AdS projection operators\cite{FukSpinor}
	\begin{equation}\label{6-21}
		\mathcal{P}^{\pm} = \frac{1\pm \frac{\tau^A}{\rho}\gamma_A}{2} =: \frac{1\pm \slashed{\tau}}{2}
	\end{equation}
	where $ \tau $ is the tangency point map, satisfying $ \tau_A\tau^A = 1 $.
	One can check that this is indeed a projection:
	\begin{equation}\label{6-22}
		\mathcal{P}^{+}\mathcal{P}^{+} =\frac{1+ \slashed{\tau}}{2}\frac{1+ \slashed{\tau}}{2} = \frac{1+2\slashed{\tau} + \slashed{\tau}\slashed{\tau}}{4} = \frac{1+ \slashed{\tau}}{2} = \mathcal{P}^{+}
	\end{equation} and similarly for others.

\section{Levi-Civita identities}\label{LCConv}
	The totally antisymmetric \textit{Levi-Civita symbol} in $ N $ dimensions is given abstractly as $ \epsilon(\pi) = \text{sign}(\pi) $, where $ \pi $ is a permutation of a set of $ N $ elements. In practice, this means for example in $ 4D $ that $ \epsilon_{1234} = 1 $ on the set $ \{1,2,3,4\} $ and the remainder of the symbol is determined by complete antisymmetry and thus cyclicity.\\
	In the course of this thesis, I use several identities which I will give in a suggestive form each.\cite{MTW,Riley2006,Lip2001}\\
	\begin{align}\label{6-23}
		\epsilon_{a_1\cdots a_k b_{k+1}\cdots b_{N}}\epsilon^{a_1\cdots a_k c_{k+1}\cdots c_N} = k!\,  \delta^{c_{k+1}\cdots c_N}_{b_{k+1}\cdots b_{N}}
	\end{align}
	with the generalised Kronecker delta given as the determinant of the matrix with entries $M^i_j = \delta^{c_i}_{b_j} $ or 
	\begin{align}\label{6-24}
		\delta^{c_{1}\cdots c_m}_{b_{1}\cdots b_{m}} = \sum_{\pi \in S_m} \text{sign}(\pi) \delta^{c_1}_{b_\pi(1)} \cdots\delta^{c_m}_{b_\pi(m)}
	\end{align}
	For example, 
	\begin{align}\label{6-25}
		\begin{split}
		\delta^{ab}_{cd} &= \delta^a_c \delta^b_d - \delta^a_d \delta^b_c\\
		\delta^{rsk}_{abc} &= \delta^r_a \delta^{sk}_{bc} + \delta^r_b \delta^{sk}_{ca} + \delta^r_c \delta^{sk}_{ab}\\
		= \delta^r_a \delta^s_b \delta^k_c &+ \delta^r_b \delta^s_c \delta^k_a + \delta^r_c \delta^s_a \delta^k_b - \delta^r_a \delta^s_c \delta^k_b - \delta^r_b \delta^s_a \delta^k_c - \delta^r_c \delta^s_b \delta^k_a
		\end{split}
	\end{align}
	In particular, one gets from (\ref{6-23})
	\begin{align}\label{6-26}
		\begin{split}
		\epsilon_{abcd}\epsilon^{rskd} &=\delta^{rsk}_{abc}\\
		\epsilon_{abcd}\epsilon^{rscd} &= 2!\,  \delta^{rs}_{ab} = 2(\delta^a_c \delta^b_d - \delta^a_d \delta^b_c)\\
		\epsilon_{abcd}\epsilon^{rbcd} &=3! \delta^r_a\\	
		\epsilon_{abcd}\epsilon^{abcd} &= 4!
		\end{split}
	\end{align}and similarly for the 5D Levi-Civita symbol.
	Determinants. Be $ A = (A^a_b)^a_b $ an $ N\times N $ matrix.
	\begin{align}\label{6-27}
		\epsilon^{b_1\cdots b_N} A^{a_1}_{b_1}\cdots A^{a_N}_{b_N} = \text{det}(A) \epsilon^{a_1\cdots a_N}
	\end{align}
	In particular,
	\begin{align}\label{6-28}
		\text{det}(A) = \frac{1}{N!} \epsilon_{a_1\cdots a_N}\epsilon^{b_1\cdots b_N} A^{a_1}_{b_1}\cdots A^{a_N}_{b_N}
	\end{align} 

\section{The internal hodge dual}\label{Hodge}
	Here, I discuss a few properties of the internal hodge dual defined, in a basis $ \{e_A\}, A\in \{1,2,3,4,5\} $, as
	\begin{align}\label{6-29}
		\circledast : \Lambda^p(\mathbb{R}^{2,3}) &\to \Lambda^{4-p}(\mathbb{R}^{2,3})\\
		\frac{1}{p!}V^{A_1\cdots A_p}e_{A_1}\cdots \wedge e_{A_p} \mapsto \frac{1}{(4-p)! p!}&\epsilon^{A_1 \cdots A_{4-p}}_{\qquad\;\;\;\; B_1\cdots B_p E} V^{B_1\cdots B_p} \tau^E \; e_{A_1}\cdots \wedge e_{A_{4-p}}
	\end{align}
	where indices are raised with the standard metric $ \eta^{AB} $ and $ \tau_A $ fulfils $ \tau_A\tau^A = 1 $. For $ p=5 $ there is no obvious way to define it, which will make sense given the examples later.\\
	Consider first, the cases $ p=1,2 $, which are relevant to this thesis. They have the expressions 
	\begin{align}\label{6-30}
		(\circledast V)_{ABC} &= \epsilon_{ABCDE} V^D \tau^E \\
		(\circledast F)_{AB} &= \epsilon_{ABCDE} F^{CD} \tau^E
	\end{align} 
	respectively. One immediately notices that $ (\circledast V)_{ABC} \tau^A = 0 = (\circledast F)_{AB} \tau^A $ , so that the duals are orthogonal to $ \tau^A $. Furthermore, one can take the double dual:
	\begin{align}\label{6-31}
		(\circledast^2 V)^{A} &= -(V^A - (\tau_CV^C)\, \tau^A )\\
		(\circledast^2 F)^{AB} &= F^{AB} + \tau^A F^{BE}\tau_E - \tau^B F^{AE}\tau_E
	\end{align} 
	One can see that a part of the expression is always the original component. Indeed, if $ V^A\tau_A =0 = F^{AE}\tau_E $, then this reduces to $ (\circledast^2 V)^{A} = -V^A $ and $ (\circledast^2 F)^{AB} = F^{AB} $, just like a dual should. As such, one can see the internal star as a dual on a subspace of the exterior algebra defined through $ {V^{A_1\cdots A_p} \tau_{A_p} = 0} $. On the complement of the algebra under this condition, the double dual vanishes, as can be seen from $ V^A = a \, \tau^A $:
	\begin{align}\label{6-32}
		(\circledast^2 V)^{A} &= -(a \tau^A - (a)\, \tau^A ) = 0
	\end{align}
	One can get an intuition behind this operation from a simple, three-dimensional analogue. Namely, consider a sphere bundle, as a subbundle of an $ \mathbb{R}^3 $-bundle, over some manifold. Take $ \tau $ to be a section of the sphere bundle, realised as a vector field which satisfies $ \tau^i\tau_i = 1$ everywhere. This can be visualised as a unit sphere embedded around the origin of $ \mathbb{R}^3 $ and a vector connecting the origin and some point on it. That vector will be dependent on the manifold point and move around smoothly on the sphere as one changes it. \\
	With this setting, the internal hodge dual of a vector field $ V $, a section of the $ \mathbb{R}^3 $-bundle, will be given by 
	\begin{align}\label{6-33}
		(\circledast V)_{i} &= \epsilon_{ijk} V^j \tau^k \\
		(\circledast^2 V)^{i} &= -(V^i - (\tau\cdot V)\, \tau^i )
	\end{align}
	Obviously, this is just the vector cross product with $ \tau $. One can thus see that the double dual of $ V $ is just its component tangential to the sphere. More generally, the double dual is the part of the form in question that is orthogonal to the tangency map $ \tau $ with respect to the ambient metric. In the case of the sphere, this tangency vector is everywhere orthogonal to the surface, which of course depends on the embedding of the spheres into the ambient spaces. While the analogy is not exact, one may think of the double dual of a form to be, just as well, the part tangential to the manifold one is describing with $\tau$. In the relevant case of the thesis, this would be the part of the forms tangential to $ AdS_4 $. As such, it is the part that can be thought of to only consist of forms on the AdS-space. This construction thus "emulates" a Hodge dual on $ AdS_4 $ through the ambient space. As a word of caution, though, one needs to remember that this construction is highly dependent on the way the internal spaces are embedded into their ambient ones. A more careful consideration is needed to decide if the properties of this dual are independent of it, and in which way. However, this is not in the scope of this thesis and will not be done here. \\
	This section is closed with some useful formulas.
	\begin{align}\label{6-34}
		\begin{split}
		\circledast(D\tau)_{abc} &= \frac{-\rho}{3!} \epsilon_{abcd} \theta^d\\
		\circledast(D\tau\wedge D\tau)_{ab} &= \frac{\rho}{2!} \epsilon_{abcd} \theta^c\wedge\theta^d\\
		\circledast(D\tau\wedge D\tau \wedge D\tau)_{a} &= \frac{-\rho}{1!} \epsilon_{abcd} \theta^b\wedge\theta^c\wedge\theta^d \\
		\circledast(D\tau\wedge D\tau\wedge D\tau\wedge D\tau) &= 4!\rho \, vol
		\end{split}
	\end{align}

  \bibliographystyle{unsrtnat}
  \bibliography{References}  

\cleardoublepage
\vspace*{2cm}
{\LARGE \sffamily{\textbf{Declaration of Academic Integrity}}}\\
\vspace*{0.2cm}\\
Hereby, I declare that I have written this thesis independently on my own and without any other sources than the ones mentioned.
\vspace*{3cm}\\
\noindent
\rule[0.5ex]{25em}{0.3pt}\\
Simon Langenscheidt \qquad \qquad Munich, \today

\end{document}